\documentclass[fleqn,usenatbib]{mnras}

\usepackage{newtxtext,newtxmath}
\usepackage[T1]{fontenc}

\DeclareRobustCommand{\VAN}[3]{#2}
\let\VANthebibliography\thebibliography
\def\thebibliography{\DeclareRobustCommand{\VAN}[3]{##3}\VANthebibliography}

\usepackage{graphicx}	% Including figure files
\usepackage{amsmath}	% Advanced maths commands
\DeclareRobustCommand{\ion}[2]{\textup{#1\,\textsc{\lowercase{#2}}}}

\newcommand{\dsb}{AT~2022dsb}
\newcommand{\erasst}{eRASSt~J154221.6-224012}

\newcommand{\namehost}{ESO~583-G004}

\newcommand{\ntdeswithearlyxray}{six}

\newcommand{\fxobsdrop}{39} % 10^(-12.46) / 10^(-14.05)

\newcommand{\dateerodiscovery}{2022-02-17}
\newcommand{\mjdpeak}{$59640.9 ^{+0.5}_{-0.4}$}
\newcommand{\phaseerassfivenoerr}{$-13.5$}

\newcommand{\specndaysafterpeak}{140}

\newcommand{\unitflux}{\text{erg\,cm}$^{-2}$\,\text{s}$^{-1}$}

\newcommand{\mstarsedfitsixtyeight}{$M_{\star}=7 ^{+4}_{-3} \times 10^{10}\,\rm{M_{\odot}}$} %

\newcommand{\mbhsedfitlogonesixtyeight}{$\log [M_{\mathrm{BH}} / M_{\odot}]=7.3 ^{+0.2}_{-0.3}$}

\newcommand{\lumerassfive}{$2.5^{+0.6}_{-0.5} \times 10^{43}$}

\newcommand{\lcofwhm}{$10400 \pm 400$}
\newcommand{\lcocentroid}{$6530 \pm 7$}
\newcommand{\lcovelocityoffset}{$-1600 \pm 300$}
\newcommand{\nttfwhm}{$10500 \pm 300$}
\newcommand{\nttvelocityoffset}{$200 \pm 400$}

\newcommand{\hstlymanalphavelocity}{3000}

\title[Early transient X-rays in a TDE]{Transient fading X-ray emission detected during the optical rise of a tidal disruption event}%
\author[Adam Malyali]{A. Malyali$^{1}$\thanks{E-mail: amalyali@mpe.mpg.de}, A. Rau$^{1}$, C. Bonnerot$^{2}$, A.~J.~Goodwin$^{3}$, Z. Liu$^{1}$, G.~E.~Anderson$^{3}$, J. Brink$^{4,5}$, \newauthor D.~A.~H. Buckley$^{4,5,6}$, A. Merloni$^{1}$, J.~C.~A.~Miller-Jones$^{3}$, I. Grotova$^{1}$, A.~Kawka$^{3}$ % Alphabetical author ordering for now!
\\
$^{1}$Max-Planck-Institut f\"ur extraterrestrische Physik,  Giessenbachstrasse 1, 85748 Garching, Germany\\
$^{2}$School of Physics and Astronomy \& Institute for Gravitational Wave Astronomy, University of Birmingham, Birmingham B15 2TT, UK\\
$^{3}$International Centre for Radio Astronomy Research, Curtin University, GPO Box U1987, Perth, WA 6845, Australia\\
$^{4}$South African Astronomical Observatory, PO Box 9, Observatory Rd, 7935 Observatory, Cape Town, South Africa\\
$^{5}$Department of Astronomy, University of Cape Town, Private Bag X3, Rondebosch 7701, South Africa\\
$^{6}$Department of Physics, University of the Free State, PO Box 339, Bloemfontein 9300, South Africa
}

\date{Accepted XXX. Received YYY; in original form ZZZ}

\pubyear{2023}

\begin{document}
\label{firstpage}
\pagerange{\pageref{firstpage}--\pageref{lastpage}}
\maketitle

\begin{abstract}
We report on the \textit{SRG}/eROSITA detection of ultra-soft ($kT=47^{+5}_{-5}$~eV) X-ray emission ($L_{\mathrm{X}}=$\lumerassfive ~erg~s $^{-1}$) from the tidal disruption event (TDE) candidate AT 2022dsb $\sim$14 days before peak optical brightness. As the optical luminosity increases after the eROSITA detection, then the 0.2--2~keV observed flux decays, decreasing by a factor of $\sim$\fxobsdrop \, over the 19 days after the initial X-ray detection.  Multi-epoch optical spectroscopic follow-up observations reveal transient broad Balmer emission lines and a broad \ion{He}{II}~4686{\AA} emission complex with respect to the pre-outburst spectrum. Despite the early drop in the observed X-ray flux, the \ion{He}{II}~4686{\AA} complex is still detected for $\sim$40 days after the optical peak, suggesting the persistence of an obscured, hard ionising source in the system. Three outflow signatures are also detected at early times: i) blueshifted H$\alpha$ emission lines in a pre-peak optical spectrum, ii) transient radio emission, and iii) blueshifted Ly$\alpha$ absorption lines. The joint evolution of this early-time X-ray emission, the \ion{He}{II}~4686{\AA} complex and these outflow signatures suggests that the X-ray emitting disc (formed promptly in this TDE) is still present after optical peak, but may have been enshrouded by optically thick debris, leading to the X-ray faintness in the months after the disruption. If the observed early-time properties in this TDE are not unique to this system, then other TDEs may also be X-ray bright at early times and become X-ray faint upon being veiled by debris launched shortly after the onset of circularisation.
\end{abstract}


%%%%%%%%%%%%%%%%%%%%%%%%%%%%%%%%%%%%%%%%%%%%%%%%%%

%%%%%%%%%%%%%%%%% BODY OF PAPER %%%%%%%%%%%%%%%%%%

\section{Introduction}
The number of stellar tidal disruption event (TDE) candidates identified in recent years has greatly increased, largely fuelled by the increasing number of wide-field, high-cadence time-domain surveys operating across the electromagnetic spectrum. Although early theoretical work predicted TDEs to produce large amplitude, ultra-soft X-ray flares originating from the centres of galaxies \citep{rees_tidal_1988} -- consistent with the first TDE candidates identified by \textit{ROSAT} \citep{trumper_rosat_1982} in the 1990s \citep{bade_detection_1996,grupe_rx_1999,komossa_discovery_1999,komossa_giant_1999,greiner_rx_2000} -- the majority of optically-selected TDE candidates do not show transient X-ray emission \citep{van_velzen_seventeen_2021,hammerstein_final_2023}. To explain the dearth of X-rays in these systems, it has been suggested that the optical emission is produced by the debris circularisation process instead of accretion (stream-stream collisions; \citealt{piran_disk_2015,shiokawa_general_2015}), or that a large fraction of the X-ray emission is reprocessed to optical/UV bands by debris enveloping the nascent disc \citep{loeb_optical_1997,lodato_multiband_2011,miller_disk_2015,metzger_bright_2016,dai_unified_2018,lu_self-intersection_2020}.

Optical spectroscopic follow-up of optically bright TDEs has led to the classifications of TDEs into different spectral types (e.g. \citealt{arcavi_continuum_2014,leloudas_spectral_2019,van_velzen_seventeen_2021}), depending on the emission lines seen in the spectra. These are i) `H’, which show transient broad Balmer emission lines, ii) `H+He’, showing transient broad Balmer emission lines and a broad emission complex around \ion{He}{II} 4686A, and iii) `He’, which show a transient broad \ion{He}{II} 4686A emission feature but no Balmer emission. An additional spectral TDE class not common in recent optically-selected TDE samples are the extreme coronal line emitters (ECLEs; \citealt{komossa_ntt_2009,wang_transient_2011,wang_extreme_2012}), which show strong emission from high-ionisation coronal lines with respect to their narrow [\ion{O}{III}]~5007~{\AA} emission.
A hard ionising source (photons with energy above 54 eV) is needed to produce the \ion{He}{II} emission seen in `He’ and `H+He’ TDEs (herein collectively referred to as He-TDEs), yet the majority of TDE candidates even in these classes do not show transient X-ray emission \citep{hammerstein_final_2023}. As it is thought that these hard photons originate from the high-energy tail of the newly formed disc, then the combination of the X-ray faintness and the \ion{He}{II} emission in these systems has been suggested as evidence for `obscured accretion’ \citep{leloudas_spectral_2019}, where an accretion disc has formed in these systems, but its high-energy emission gets reprocessed into the optical band by an optically-thick gaseous envelope. Several TDE candidates have also shown broad \ion{He}{II} lines close to peak optical brightness \citep{blagorodnova_iptf16fnl_2017,nicholl_outflow_2020,wevers_elliptical_2022}, which under the assumption of an obscured accretion-driven origin, suggests efficient circularisation of the debris into a disc post-disruption. 

Here, we report on multi-wavelength observations of the TDE candidate AT~2022dsb, which shows a factor of \fxobsdrop~decrease in its 0.2--2~keV observed flux during the optical rise. Section~\ref{sec:discovery} describes the discovery of \dsb, whilst sections~\ref{sec:observations} and ~\ref{sec:data_analysis} detail multi-wavelength observations of the system and their analysis, respectively. In section~\ref{sec:early_xray_emission_tdes}, we review previous X-ray observations of TDEs at early times and compare these with \dsb. The implications of our observational campaign are discussed in  section~\ref{sec:discussion}, and our conclusions in section~\ref{sec:conclusions}. All magnitudes are reported in the AB system and corrected for Galactic extinction using $A_{\mathrm{V}}=0.62$~mag, obtained from \citet{schlafly_measuring_2011},  $R_{\mathrm{V}}=3.1$ and a Cardelli extinction law \citep{cardelli_relationship_1989}. The effective wavelength for each filter was retrieved from the SVO Filter Profile Service\footnote{\url{http://svo2.cab.inta-csic.es/theory/fps/}}. All dates and times will be reported in universal time (UT). 

\section{Discovery}\label{sec:discovery} %eRASSt_J154221.6-224012.1
\begin{figure}
    \centering
\includegraphics{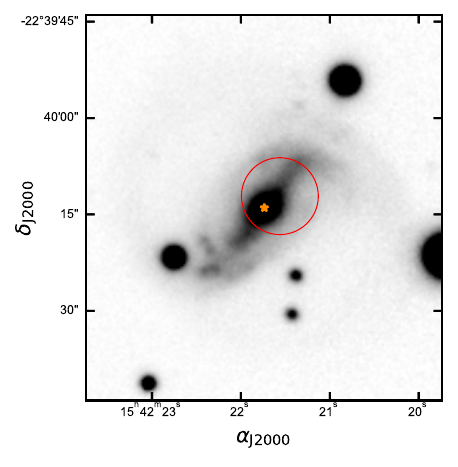}
    \caption{Finder chart for AT~2022dsb (DESI LS DR10 $g$-band image). The red circle denotes the 3$\sigma$ uncertainty on the eROSITA source position in eRASS5, whilst the dark orange star marks the \textit{Gaia}~DR3 optical centre of the host galaxy \namehost.}
    \label{fig:finder_chart}
\end{figure}

\dsb/ \erasst\, was independently discovered by the extended ROentgen Survey with an Imaging Telescope
Array (eROSITA; \citealt{predehl_erosita_2021}), the soft X-ray instrument on board the \textit{Spektrum-Roentgen-Gamma} (SRG; \citealt{sunyaev_srg_2021}) observatory,  during a systematic search for TDE candidates in the fifth eROSITA All-Sky Survey (eRASS5), when it was observed on \dateerodiscovery\, as a new, bright (0.2-2~keV observed flux of $\sim 3\times 10^{-13}$~erg~cm$^{-2}$~s$^{-1}$), ultra-soft (section~\ref{sec:spec_fitting}) X-ray point source. Using the \textit{eROSITA Science Analysis Software} pipeline (eSASS\footnote{Version: eSASSusers\_211214.}; \citealt{brunner_erosita_2022}), the source was localised to (RA$_\mathrm{J2000}$, Dec$_\mathrm{J2000})$=(15h42m21.6s,-22$^{\circ}$40$^\prime$12.1$^{\prime\prime}$), with a 1$\sigma$ positional uncertainty of 1.9$^{\prime \prime}$ (68\% confidence), consistent with the galaxy \namehost\ at $z=0.0235$ (Fig.~\ref{fig:finder_chart}). 
No X-ray source had been detected within 30\arcsec of this position in any of the previous four eRASS, with a 3$\sigma$ upper limit on the 0.2-2~keV band flux of $5\times 10^{-14}$ \unitflux, assuming the same spectral model fitted to the eRASS5 spectrum \citep{liu_srgerosita_2022}. The last non-detection by eROSITA occurred $\sim$6 months before the eRASS5 detection. 

\dsb\, was later publicly classified on 2022-03-02 as a TDE candidate in the TNS report TNSCR-2022-584 \citep{fulton_epessto_2022}, after the discovery and reporting of optical transient emission (associated to the nucleus of the host galaxy \namehost) initially by ASAS-SN on 2022-03-01 in \citet{stanek_asas-sn_2022}, and then by both the Asteroid Terrestrial Impact Last Alert System (ATLAS; \citealt{tonry_atlas_2018}), and the Zwicky Transient Facility (ZTF; \citealt{bellm_zwicky_2019,graham_zwicky_2019}) by the ALeRCE alert broker \citep{forster_automatic_2021}. 
%eROSITA detection preceded all public reportings of the associated optical transient.

\subsection{Host galaxy}
A pre-outburst optical spectrum of \namehost\, was taken in 2002 during the Six-Degree Field (6dF; \citealt{jones_6df_2009}) galaxy survey. A recent analysis of the narrow emission lines in this optical spectrum classified the system as a type II AGN \citep{chen_uniformly_2022}, according to the criteria presented in \citet{kewley_theoretical_2001}. However, the pre-outburst \textit{AllWISE} \citep{wright_wide-field_2010,mainzer_initial_2014} colour of the host, $W1-W2=0.00 \pm 0.03$ mag, suggests that its mid-infrared emission is dominated by the galaxy light, instead of the luminous emission from a dusty torus surrounding an AGN \citep{stern_mid-infrared_2012,assef_wise_2018}. \namehost\, may have hosted a low-luminosity AGN prior to the outburst of \dsb, similar to other TDE candidates (e.g.~ASASSN-14li, \citealt{holoien_six_2016}; AT~2019qiz, \citealt{nicholl_outflow_2020}).

\begin{figure}
    \centering
    \includegraphics[scale=0.85]{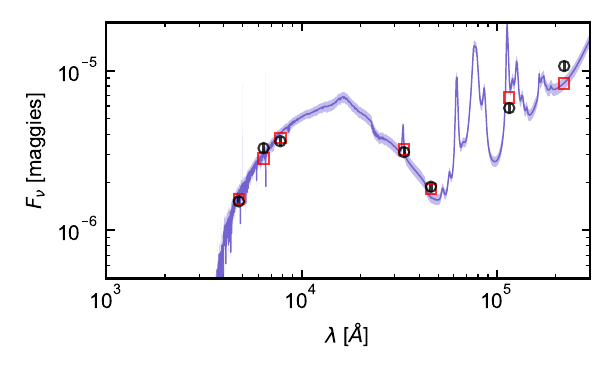}
    \caption{SED fit to the LS DR10 photometry of the host galaxy of \dsb . %($g$, $i$, $W1$, $W2$, $W3$ and $W4$ bands). 
    The observed photometry is plotted with black edged circular markers, whilst the posterior model and model photometry is shown in blue (solid line represents the median, shaded region encloses the middle 90\% of the posterior) and red, respectively. }%\textbf{TODO: update plot with latest photometry......}
    \label{fig:sed_fit}
\end{figure}

We fitted the DESI Legacy DR10 \citep{dey_overview_2019} archival photometry of the host galaxy ($g$, $i$, $W1$, $W2$, $W3$ and $W4$ bands\footnote{No $z$-band photometry was available for the host of \dsb\, in DR10, and all DR10 photometry was taken before August 2021.}) with the stellar population inference tool \texttt{Prospector} \citep{johnson_stellar_2021}, which uses a python wrapper \citep{foreman-mackey_python-fsps_2014} of the Flexible Stellar Population
Synthesis code \citep{conroy_fsps_2010} for generating the SEDs of stellar populations. The SED model includes both stellar and nebular emission, as well as dust attenuation and emission, and adopts a Chabrier initial mass function (IMF; \citealt{chabrier_galactic_2003}); the free parameters are the total stellar mass of the galaxy ($M_{\star}$, the sum of both living and remnant stars), the metallicity ($\log (Z/Z_{\odot})$), the age of the galaxy ($t_{\mathrm{age}}$), the decay timescale under an exponentially declining star formation model ($\tau_{\mathrm{SF}}$), and the host galaxy dust extinction ($A_{\mathrm{V}}$). Posterior distributions were sampled using the dynamic nested sampler \citep{skilling_nested_2004,skilling_nested_2006,higson_dynamic_2019} \texttt{dynesty} \citep{speagle_dynesty_2020,koposov_joshspeagledynesty_2023}, with the posterior model shown in Fig.~\ref{fig:sed_fit} and the parameter estimates in Table~\ref{tab:sed_fit_parameters}. From the inferred \mstarsedfitsixtyeight
and using the relation between $M_{\mathrm{BH}}$ and $M_{\star}$ in \citet{reines_relations_2015}, we infer \mbhsedfitlogonesixtyeight .  % Reines relation is between total stellar mass and bh mass
%As a consistency check, we also estimate $M_{\mathrm{BH}}\sim 4.7 \times 10^6$~$\mathrm{M}_{\odot}$ (scatter of X) when using an $M_{\star}$ inferred from the \citet{kettlety_galaxy_2018} relation between $M_{\mathrm{*}}/L_{\mathrm{W1}}=0.65 \pm 0.07$, where $L_{\mathrm{W1}}$ is the host galaxy luminosity in the \textit{WISE} $W1$-band.
%\textbf{TODO: Double check the green-valley...}
% CAPTION: Host galaxy properties inferred via SED fitting to the LS DR10 archival photometry, with CR denoting the credible region for a parameter.
\begin{table}
\centering
\caption{Host galaxy properties inferred via SED fitting to the archival photometry, with CR denoting the credible region for a parameter.}
\label{tab:sed_fit_parameters}
\begin{tabular}{cc}
\hline
Parameter & 68\% CR \\
\hline
$M_{\mathrm{\star}}$$/ 10^{10} M_{\odot}$ & $6.9 ^{+3.9}_{-3.3}$  \\
$\log [Z/ \mathrm{Z}_{\odot}]$ & $-0.7 ^{+0.3}_{-0.3}$ \\
$A_{\mathrm{V}}$/mag & $0.62 ^{+0.05}_{-0.06}$ \\
$t_{\mathrm{age}}$/Gyr & $5.4 ^{+4.3}_{-3.2}$ \\
$\tau _{\mathrm{SF}}$/Gyr & $0.2 ^{+0.5}_{-0.1}$ \\
\end{tabular}
\end{table}

\section{Observations and data reduction}\label{sec:observations}
\begin{figure*}
    \centering
    \includegraphics[scale=0.8]{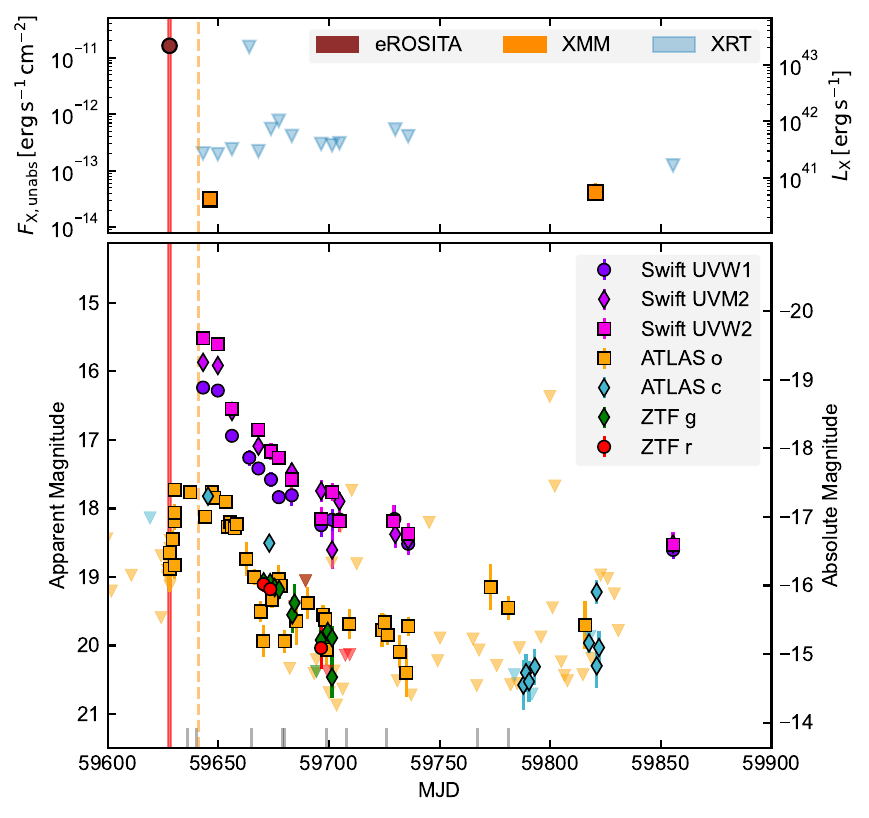}
    \caption{0.2--2~keV X-ray (top) and optical-UV (bottom) evolution of \dsb. Solid markers denote $>3\sigma$ detections in each epoch, whereas translucent triangles mark $3\sigma$ upper limits. The vertical red line marks the eROSITA observation of \dsb , whilst the vertical orange line marks the inferred time of optical peak on MJD 59641 (section~\ref{sec:analysis_photometric}), $\sim 14$ days after the eRASS5 detection. %To highlight the evolution of the observed X-ray emission here, the X-ray fluxes in the top panel have not been corrected for Galactic absorption (corrected values are presented in Table~\ref{tab:x_ray_lightcurve_table}).
    }
    \label{fig:multiwavelength_lc}
\end{figure*}

\begin{figure}
    \centering
    \includegraphics[scale=0.55]{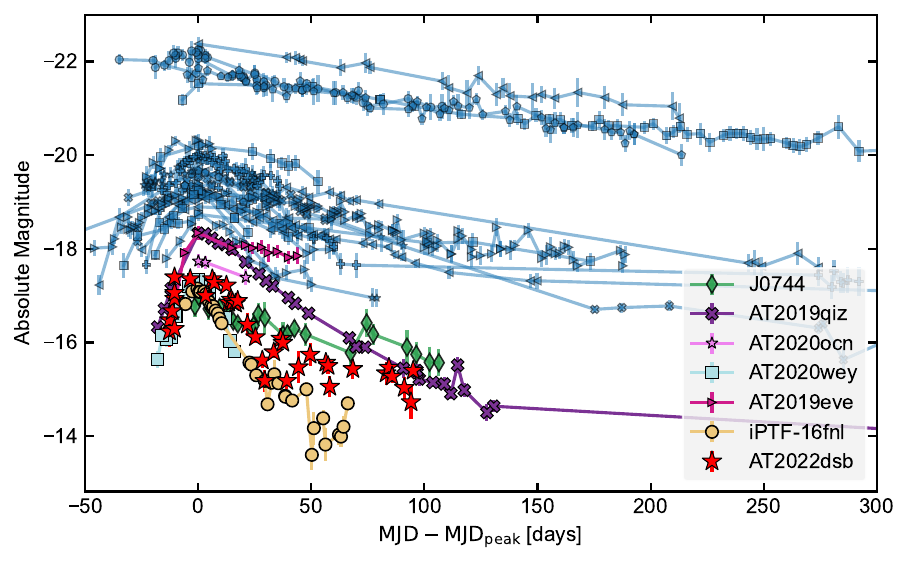}
    \caption{Comparison of the ATLAS $o$-band lightcurve of \dsb\, (red stars) with the $g$-band lightcurves of ZTF-selected TDEs (blue markers) reported in \citet{hammerstein_final_2023}. TDEs of similar peak absolute magnitudes are highlighted in non-dark blue colours, and we include data from the `faint and fast' TDE iPTF-16fnl \citep{blagorodnova_iptf16fnl_2017}, and the `faint and slow' TDE eRASSt~J074426.3+291606 (the faintest optically-bright TDE observed to date, J0744; \citealt{malyali_erasst_2023}).}
    \label{fig:optical_tde_lc_comparison}
\end{figure}

After the initial eROSITA discovery, additional X-ray observations of \dsb\, were obtained with \textit{XMM-Newton} (section~\ref{sec:obs_xray_xmm}) and \textit{Swift} XRT (section~\ref{sec:obs_xray_xrt}); %, \textit{NICER} (section~\ref{sec:obs_xray_nicer})
 collectively, these sample the X-ray emission from a TDE pre-peak, near-peak and post-peak optical brightness. The optical, UV and radio evolution was also monitored with ground-based photometry (section~\ref{sec:obs_phot_ground}), \textit{Swift} UVOT (section~\ref{sec:obs_phot_uvot}) and the Australia Telescope Compact Array (ATCA; section~\ref{sec:obs_radio}), respectively. The full multi-wavelength lightcurve of \dsb\, is depicted in Fig.~\ref{fig:multiwavelength_lc}, and a comparison of the optical lightcurve with other TDE candidates is shown in Fig.~\ref{fig:optical_tde_lc_comparison}. A log of all X-ray observations and the inferred fluxes is presented in Table~\ref{tab:x_ray_lightcurve_table}, whilst the optical and UV photometry can be found in Table~\ref{tab:uvot_photometry}.
\begin{table*}
\centering
\caption{X-ray lightcurve of AT~2022dsb. $F_{\rm 0.2-2keV, obs}$ and $F_{\rm 0.2-2keV, unabs}$ are the observed (not corrected for Galactic absorption) and unabsorbed 0.2--2~keV band fluxes in units of \unitflux. $\log [L_{\rm 0.2-2keV}$] is inferred from $F_{\rm 0.2-2keV, unabs}$. MJD is computed from the midpoint of MJD$_{\mathrm{start}}$ and MJD$_{\mathrm{stop}}$. The fluxes have been estimated from the best fitting model (Table~\ref{tab:x_ray_model_fits}), with the 3$\sigma$ upper limits on the count rates converted to fluxes using the best fitting model to the eRASS5 spectrum.}
\label{tab:x_ray_lightcurve_table}
\begin{tabular}{cccccccc}
\hline
MJD & MJD$_{\mathrm{start}}$ & MJD$_{\mathrm{stop}}$ & Instrument & ObsID & $\log [F_{\rm 0.2-2keV, obs}]$  & $\log [F_{\rm 0.2-2keV, unabs}$] & $\log [L_{\rm 0.2-2keV}$] \\
\hline
59627.939 & 59627.439 & 59628.439 & eROSITA & eRASS5 & $-12.46 ^{+0.07}_{-0.08}$ & $-10.79 ^{+0.10}_{-0.14}$ & $43.34 ^{+0.10}_{-0.14}$ \\
59643.115 & 59643.004 & 59643.226 & XRT & 00015054002 & $<$-12.69 & $<$-11.11 & $<$43.02 \\
59646.226 & 59646.088 & 59646.363 & EPIC PN & XMM1 & $-14.05 ^{+0.05}_{-0.06}$ & $-13.50 ^{+0.10}_{-0.10}$ & $40.62 ^{+0.10}_{-0.10}$ \\
59649.772 & 59649.702 & 59649.842 & XRT & 00015054003 & $<$-12.71 & $<$-11.12 & $<$43.01 \\
59656.156 & 59656.150 & 59656.161 & XRT & 00015054005 & $<$-12.62 & $<$-11.03 & $<$43.10 \\
59663.969 & 59663.968 & 59663.969 & XRT & 00015054006 & $<$-10.81 & $<$-9.22 & $<$44.90 \\
59668.128 & 59668.030 & 59668.225 & XRT & 00015054007 & $<$-12.65 & $<$-11.07 & $<$43.06 \\
59673.808 & 59673.805 & 59673.811 & XRT & 00015054008 & $<$-12.26 & $<$-10.67 & $<$43.45 \\
59677.316 & 59677.178 & 59677.453 & XRT & 00015054009 & $<$-12.12 & $<$-10.53 & $<$43.60 \\
59683.209 & 59683.205 & 59683.214 & XRT & 00015054010 & $<$-12.38 & $<$-10.80 & $<$43.33 \\
59696.480 & 59696.206 & 59696.755 & XRT & 00015054011 & $<$-12.53 & $<$-10.94 & $<$43.19 \\
59701.425 & 59701.251 & 59701.598 & XRT & 00015054012 & $<$-12.55 & $<$-10.96 & $<$43.16 \\
59704.749 & 59704.513 & 59704.986 & XRT & 00015054013 & $<$-12.51 & $<$-10.93 & $<$43.20 \\
59729.976 & 59729.051 & 59730.902 & XRT & 00015054014 & $<$-12.27 & $<$-10.68 & $<$43.44 \\
59735.808 & 59735.801 & 59735.816 & XRT & 00015054015 & $<$-12.39 & $<$-10.80 & $<$43.32 \\
59820.520 & 59820.376 & 59820.664 & EPIC PN & XMM2 & $-14.18 ^{+0.08}_{-0.08}$ & $-13.38 ^{+0.16}_{-0.15}$ & $40.75 ^{+0.16}_{-0.15}$ \\
59855.517 & 59855.146 & 59855.888 & XRT & 00015054016 & $<$-12.91 & $<$-11.32 & $<$42.81 \\
\end{tabular}
\end{table*}

\subsection{X-ray}\label{sec:obs_xray}
\subsubsection{eROSITA}\label{sec:obs_xray_erosita}
\begin{figure}
    \centering
    \includegraphics[scale=0.8]{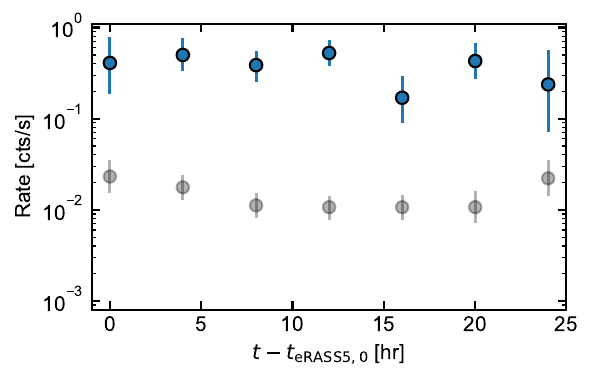}
    \caption{eRASS5 lightcurve of \dsb\, in the 0.2--2~keV band. The blue markers denote the source count rates (corrected for vignetting), whilst the grey markers show the estimated background count rates. $t-t_{\mathrm{eRASS5,0}}$ is the time relative to the start of eROSITA's observations of \dsb\ in eRASS5 (MJD=59627.439). \dsb\, is persistently bright over the day-long monitoring window in eRASS5.}
    \label{fig:xray_erass5_lightcurve}
\end{figure}

The position of AT~2022dsb was observed by eROSITA during the first four eRASS, (denoted eRASS1, eRASS2, eRASS3, and eRASS4, respectively) on 2020-02-27, 2020-08-26, 2021-02-11, and 2021-08-20. %No source was detected , with 3sigma UL from the stack X. 
During eRASS5, eROSITA first observed \dsb\, on 2022-02-17, scanning over its position seven times over the following day, with each visit separated by four hours (Fig.~\ref{fig:xray_erass5_lightcurve}). Using the eSASS task SRCTOOL, we generated source and background spectra by extracting counts from a source aperture of radius 30$^{\prime \prime}$, and a background annulus of inner and outer radii 90$^{\prime \prime}$ and 240$^{\prime \prime}$, respectively, with both apertures centred on the eROSITA position of \dsb. The same source and background apertures were used to generate a 0.2--2~keV lightcurve using SRCTOOL, shown in Fig.~\ref{fig:xray_erass5_lightcurve}. \dsb\, is clearly detected above background by eROSITA in each of the seven observations within eRASS5 (i.e. is persistently bright, instead of showing a `one-off' short flaring), providing a lower limit of 1~day on the duration of X-ray emission at early-times in \dsb. A log of X-ray observations is presented in Table~\ref{tab:x_ray_lightcurve_table}.
%erass5_MJD-OBS 2022-02-17 22:31:57.833
%erass5_MJD_START 2022-02-17 10:31:57.833
%erass5_MJD_STOP 2022-02-18 10:31:57.833

\subsubsection{XMM-Newton}\label{sec:obs_xray_xmm}%0884960701 
Additional observations of \dsb\, were performed with \textit{XMM} (P.I. Z.~Liu), with the first taking place $\sim$19 days after the eRASS5 detection on 2022-03-08, and then $\sim$173 days after this on 2022-08-29; observations were carried out in imaging mode with the medium filter. To reduce and analyse the \textit{XMM} data, we used HEASOFT (version 6.29), the \textit{XMM} Science Analysis Software (SAS) (version 20211130\_0941), and the latest calibration data files. Calibrated event files were generated from the Observation Data Files (ODF) using \texttt{emproc} and \texttt{epproc} for the MOS and PN cameras, respectively, and periods of high particle background during each observation were filtered out following the \textit{XMM} Science Operation Centre recommended procedures. This resulted in 18.0~ks and 13.3~ks exposures for the first and second observations, respectively.  Source spectra were extracted from a circle of radius 20$^{\prime \prime}$, centred on the \textit{Gaia} EDR3 \citep{gaia_collaboration_gaia_2021} position of \namehost, whilst background spectra were extracted from an annulus with inner and outer radii 76$^{\prime \prime}$ and 144$^{\prime \prime}$ respectively. Only events with \texttt{PATTERN}<=4 and \texttt{FLAG}==0 were extracted for PN, whilst PATTERN<=12 was applied for MOS1 and MOS2.

\subsubsection{Swift XRT}\label{sec:obs_xray_xrt}
AT~2022dsb was further monitored in the 0.3-10~keV band with the XRT instrument \citep{burrows_swift_2005} on-board the \textit{Neil Gehrels Swift} observatory \citep{gehrels_swift_2004}\footnote{P.I. for \textit{Swift} observations: A. Malyali, P. Charalampopoulos, J. Hinkle, I. Lypova.}. XRT observations commenced on 2022-03-05, $\sim$16 days after the eRASS5 observation, and were performed in photon counting mode. These were then analysed with the online XRT product building tool provided by the UK Swift Science Data Centre's (UKSSDC)  \citep{evans_online_2007,evans_methods_2009}. AT~2022dsb was not detected in any of the XRT observations, with 3$\sigma$ upper limits on the 0.3-2~keV count rates computed using the method presented in \citet{kraft_determination_1991}. These were then converted to 0.2-2~keV fluxes using webPIMMs\footnote{\url{https://heasarc.gsfc.nasa.gov/cgi-bin/Tools/w3pimms/w3pimms.pl}}, where we adopted the spectral model inferred from our BXA fit to the eRASS5 spectrum.

%\subsubsection{NICER}\label{sec:obs_xray_nicer}
%\textit{\begin{itemize}
%    \item Monitoring campaign roughly coincides with the XRT campaign, and is likely to represent a series of non-detections (i.e. although not properly analysed yet, inclusion of NICER data is not likely to significantly change the science of the paper). Double-check for sources within NICERs FoV.
%\end{itemize}}
%%%To include, or not to include?

\subsection{Photometry}\label{sec:obs_phot}
\subsubsection{Ground-based photometry}\label{sec:obs_phot_ground}
We obtained ATLAS $o$ and $c$ band \citep{tonry_atlas_2018} lightcurves of AT~2022dsb using the online forced photometry server %\footnote{\url{https://fallingstar-data.com/forcedphot/}} 
\citep{smith_design_2020,shingles_release_2021}. For late-time observations (MJD$>$59635), we performed a weighted rebin of the lightcurve into 1 day intervals. To improve the sampling of the lightcurve around peak optical brightness, then no such rebinning was performed for observations performed during the optical rise and the early part of the decay ($59620<\mathrm{MJD}<59635$). To remove epochs of low quality photometry in the ATLAS lightcurve, we discarded datapoints where the semi-major axis of the fitted PSF model was greater than 3~pixels (1.86$^{\prime \prime}$ per pixel).

In addition, $g$ and $r$-band lightcurves\footnote{This is generated from science images that have already been reference image subtracted} were generated using the ZTF forced photometry service \citep{masci_zwicky_2019}, which were then calibrated using the method developed by Miller et al. (in prep.) for the ZTF Bright Transient Survey\footnote{\url{https://github.com/BrightTransientSurvey/ztf_forced_phot}}. No significant optical variability is seen in the ZTF lightcurves before the 2022 outburst, and we note that the ZTF observations do not sample the rise and peak optical brightness (Fig.~\ref{fig:multiwavelength_lc}).

\subsubsection{\textit{Swift} UVOT}\label{sec:obs_phot_uvot}
%\textit{Swift}  observations%\footnote{PI: Malyali, Charalampopoulos, Lypova, Hinkle}
Over the course of the \textit{Swift} monitoring campaign, AT~2022dsb was observed by the UVOT \citep{roming_swift_2005} instrument across all filters ($V$, $B$, $U$, $UVW1$, $UVM2$ and $UVW2$), although the number of filters used varied between each observation (see photometry in Table~\ref{tab:uvot_photometry}). In this work, we use observations performed only in the $UVW1$, $UVM2$ and $UVW2$ filters, since the lightcurve sampling is highest in these bands, and the optical coverage is already provided by ATLAS and ZTF. We first downloaded the level 2 UVOT sky images from the UK Swift Science Data Centre, before computing aperture photometry on these with the \texttt{uvotsource} task (HEASOFT v6.29, CALDB v20201215), using a 5$^{\prime \prime}$ radius source aperture, and a nearby, source-free circular aperture of of radius 15$^{\prime \prime}$ for the background. Lastly, the recommended Small Scale Sensitivity check\footnote{\url{https://swift.gsfc.nasa.gov/analysis/uvot_digest/sss_check.html}} was completed.

\subsection{Optical spectroscopy}\label{sec:obs_spec}
\begin{figure}
    \centering
\includegraphics[scale=0.9]{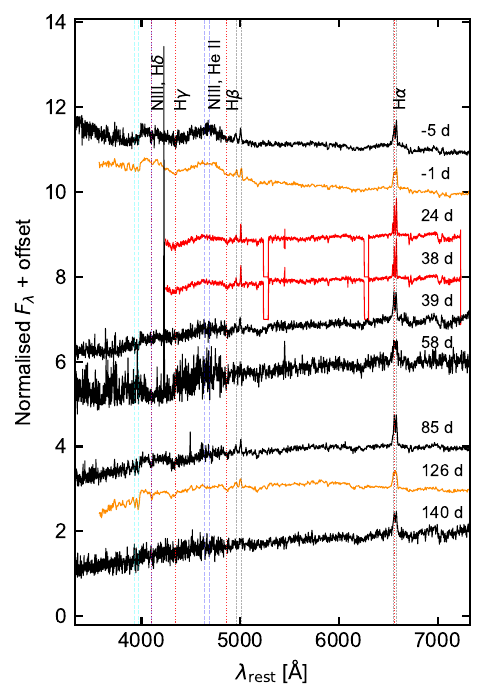}
    \caption{Optical spectroscopic evolution of AT~2022dsb. The phase of the observation with respect to the inferred optical peak (MJD=\mjdpeak) is shown on the right hand side above each spectrum. Black, orange and red spectra were obtained using LCO/FLOYDS, NTT/EFOSC2 and SALT/RSS, respectively. The archival spectrum of the host galaxy is presented in Fig.~\ref{fig:archival_optical_spectrum}.}
    \label{fig:optical_spectroscopic_evolution}
\end{figure}

The first follow-up optical spectrum of AT~2022dsb was obtained on 2022-02-26 (MJD$=59636$, $\sim$5 days before optical peak), using the FLOYDS spectrograph mounted on the 2m Las Cumbres Observatory (LCO; \citealt{brown_cumbres_2013})\footnote{Proposal ID CON2022A-001, PI: M. Salvato.} telescope at Haleakala Observatory. Data processing and spectrum extraction were performed by the automatic FLOYDS pipeline at LCO (further details on the spectroscopic data reduction are presented in section~\ref{sec:appendix_spec_reduction}). This spectrum (Fig.~\ref{fig:optical_spectroscopic_evolution}) shows transient broad Balmer emission lines (H$\alpha$, H$\beta$), a broad emission complex around 4600\AA, and a blue continuum, with the transient nature confirmed through comparison to archival and late-time optical spectra. In addition, narrow emission lines (H$\alpha$, H$\beta$, [\ion{N}{II}]~6548{\AA} and 6583~{\AA}, and the high-ionisation lines [\ion{O}{III}] 4959{\AA} and 5007~{\AA}), as well as several host galaxy absorption features, are clearly present. No strong blue continuum or broad emission lines were seen in a pre-outburst optical spectrum taken on 2002-04-15 during the 6dF Galaxy Survey (6dFGS; \citealt{jones_6df_2009}). %Unclear if the feature at 4100\AA is an emission feature or not. 
Over the \specndaysafterpeak\, days of spectroscopic monitoring of \dsb\, after optical peak, the strength of the broad emission lines and the blue continuum relative to the host galaxy decreases (Fig.~\ref{fig:optical_spectroscopic_evolution}). Zoom-in plots on the evolution of the H$\alpha$ and \ion{He}{II} complexes are presented in Fig.~\ref{fig:optical_spectroscopic_zoomins}.

\subsection{Radio}\label{sec:obs_radio}
\subsubsection{Archival}
The Karl G. Jansky Very Large Array Sky Survey (VLASS; \citealt{lacy_karl_2020}) observed the coordinates of AT2022dsb on 2020-11-03 and 2018-02-15, approximately 1 and 4 years prior to the detection of the transient event. There is no source present at the location of AT2022dsb in either of these observations, with a 3$\sigma$ upper limit of 507$\mu$Jy and 419$\mu$Jy at 3\,GHz for the 2020 and 2018 observations respectively. % https://ui.adsabs.harvard.edu/abs/2020PASP..132c5001L/abstract

\subsubsection{Follow-up}
We observed the coordinates of AT2022dsb on three occasions with the Australia Telescope Compact Array (ATCA) between 2022 March and November (project C3334, PI Anderson/Goodwin). Observations were taken in the 4-cm band with the dual 5.5 and 9\,GHz receiver. Further, more detailed radio spectral monitoring of AT2022dsb is being carried out and will be published in a follow-up paper (Goodwin et al., in prep.). Because of the early eROSITA detection of \dsb, then the ATCA observations presented here represent one of the earliest radio detections of a TDE.

\begin{figure}
    \centering
\includegraphics[width=\columnwidth]{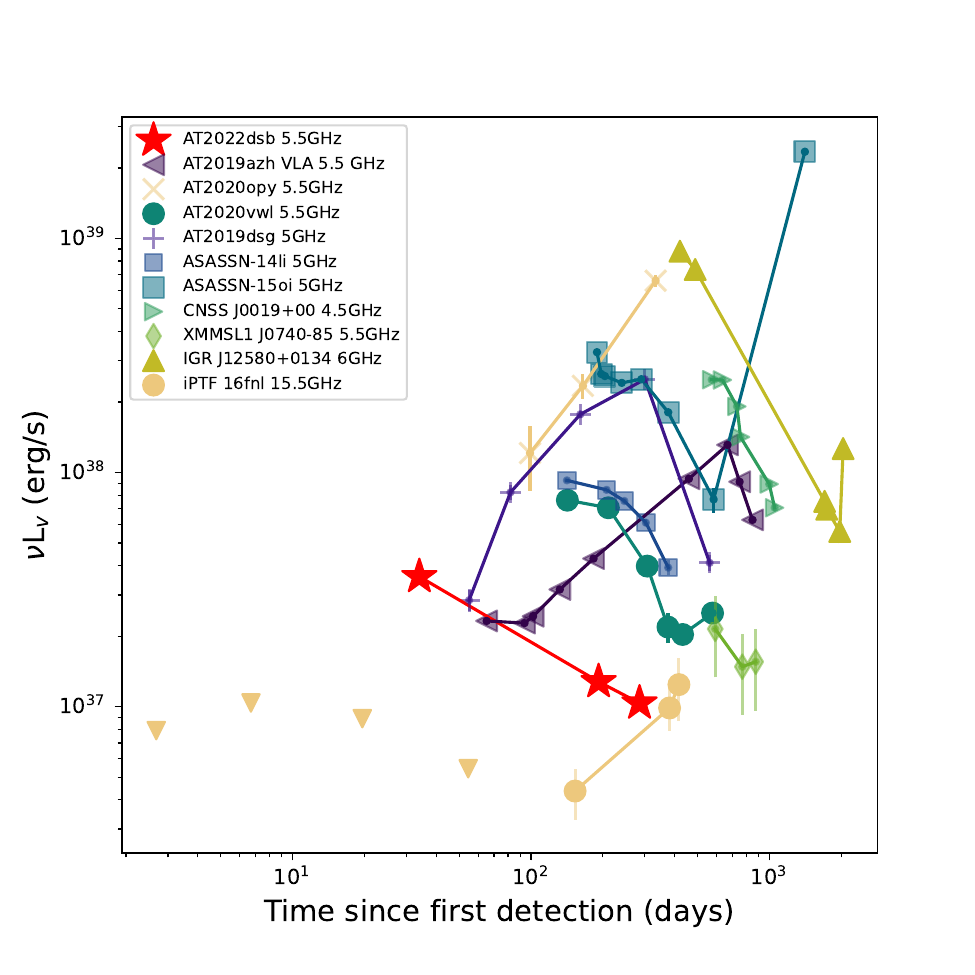}
    \caption{5.5\,GHz radio luminosity of AT2022dsb (red stars) compared to a selection of other radio-detected thermal TDEs (AT~2019azh, \citealt{goodwin_at2019azh_2022}; AT~2020opy, \citealt{goodwin_radio_2022}; AT~2019dsg, \citealt{cendes_radio_2021}; ASASSN~14li, \citealt{alexander_discovery_2016}; ASASSN~15oi, \citealt{horesh_delayed_2021}; CNSS~J0019+00, \citealt{anderson_caltech-nrao_2020}; XMMSL1~J0740-85, \citealt{alexander_radio_2017}; IGR~J12580+0134, \citealt{irwin_erratum_2018}; AT2020vwl \citealt{goodwin_radio-emitting_2023}; AT2018hyz \citealt{cendes_mildly_2022}). The horizontal axis indicates the time since first detection of the source at optical or X-ray wavelengths.}
    \label{fig:radiolc}
\end{figure}

The ATCA data were reduced using the Common Astronomy Software Application (CASA v 5.6.3; \citealt{the_casa_team_casa_2022}) using standard procedures including flux and bandpass calibration with PKS 1934-638 and phase calibration with PKS 1514-241. The target field was imaged using the CASA task \texttt{tclean} with an image size of 4000 pixels and a cellsize of 0.3\,arcsec at 5.5\,GHz and an image size of 4000 pixels and a cellsize of 0.2\,arcsec at 9\,GHz. In all observations a point source was detected at the location of AT2022dsb. The flux density of the point source was extracted in the image plane using the CASA task \texttt{imfit} by fitting an elliptical Gaussian fixed to the size of the synthesized beam. A summary of the ATCA observations is given in Table \ref{tab:ATCAobs} and the 5.5\,GHz lightcurve of AT2022dsb is plotted in Figure \ref{fig:radiolc} along with a selection of other radio-detected TDEs for comparison. Both the variability of the detected 5.5\,GHz and 9\,GHz radio emission and that the initial detection is above the 3$\sigma$ VLASS 3\,GHz upper limits years prior to the TDE suggest that the radio emission is likely related to the transient event and is not purely host galaxy emission. Although this VLASS upper limit is at a lower frequency than the ATCA observations (5.5 and 9.0 GHz), the ATCA spectrum is steep in the first epoch, so a spectral turnover would be needed in order to match the VLASS upper limit, which is not consistent with the host galaxy emission (that should be steep). % JMJ 
\begin{table}
\centering
\caption{ATCA radio observations of AT~2022dsb.}
\label{tab:ATCAobs}
\begin{tabular}{p{1cm}p{2cm}p{1cm}p{1cm}p{1.5cm}}
\hline
MJD & Date & Array config. & Frequency (GHz) & Flux density ($\mu$Jy) \\
\hline
59661 & 2022-03-23 & 6A & 5.5 & $593\pm19$ \\
59661 & 2022-03-23 & 6A & 9 & $536\pm17$ \\
59819 & 2022-08-28 & 6D & 5.5 & 211$\pm$11 \\
59819 & 2022-08-28 & 6D & 9 & 152$\pm$10 \\
59912 & 2022-11-29 & 6C & 5.5 &  171$\pm$8 \\
59912 & 2022-11-29 & 6C & 9 &  127$\pm$6 \\
\hline
\end{tabular}
\end{table}

\section{Data analysis}\label{sec:data_analysis}
\subsection{X-ray spectral fitting}\label{sec:spec_fitting}
The X-ray spectra were analysed using the Bayesian X-ray Analysis software (BXA; \citealt{buchner_x-ray_2014}), which connects the nested sampling algorithm UltraNest\footnote{\url{https://johannesbuchner.github.io/UltraNest/}} \citep{buchner_ultranest_2021}
with the fitting environment CIAO/Sherpa \citep{fruscione_ciao_2006}. The eROSITA and \textit{XMM} PN spectra were fitted in the 0.2--8~keV and 0.2--10~keV range, respectively. A joint fit of the source and background spectra was performed, using the C-statistic for fitting \citep{cash_generation_1976}, and modelling the background using the principal component analysis (PCA) technique described in \citet{simmonds_xz_2018}. The Galactic absorption is modelled with a total (HI and H$_2$) Galactic hydrogen column density of $1.73\times10^{21}$~cm$^{-2}$ \citep{willingale_calibration_2013}, cosmic abundances from \citet{wilms_absorption_2000} and cross sections from \citet{verner_atomic_1996}.

Each of the eROSITA and \textit{XMM} PN spectra were fitted with the following source models, commonly used to fit the X-ray spectra of TDEs: i) \texttt{zbbody}: redshifted blackbody ii) \texttt{zpowerlaw}: redshifted power-law, iii) \texttt{zbremsstrahlung}: redshifted thermal bremsstrahlung. %iv) \texttt{ztbabs*zbbody}: an absorbed blackbody, and v) \texttt{ztbabs*zpowerlaw}: an absorbed powerlaw, with absorption occurring in the host galaxy of \dsb\, for models iv) and v). 
%As both the eROSITA and PN source spectra contain relatively low numbers of counts (X, Y, Y respectively,), only a limited number of simple spectral models were considered here. 
To assess the goodness of fit and compare between different fitted models, we use the Akaike Information Criterion ($AIC$), defined as $AIC=2k - 2 \ln{\hat{\mathcal{L}}}$, where $k$ is the number of free-parameters in the fitted model, and $\hat{\mathcal{L}}$ the estimated maximum likelihood from the spectrum fitting; the lower the value of the AIC, the better the fit to the spectrum. An overview of the spectral fit parameters are listed in Table~\ref{tab:x_ray_model_fits}.

\begin{figure}
    \centering
    \includegraphics[scale=0.8]{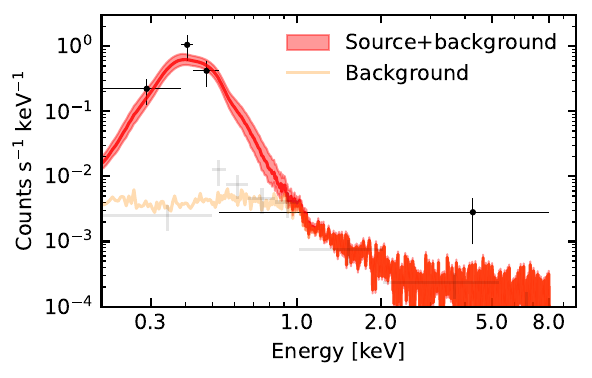}
    \includegraphics[scale=0.8]{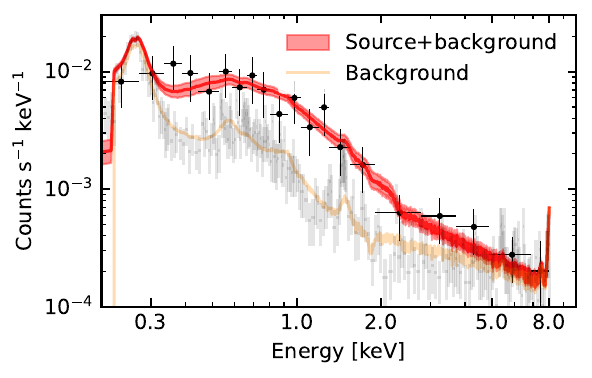}
    \includegraphics[scale=0.8]{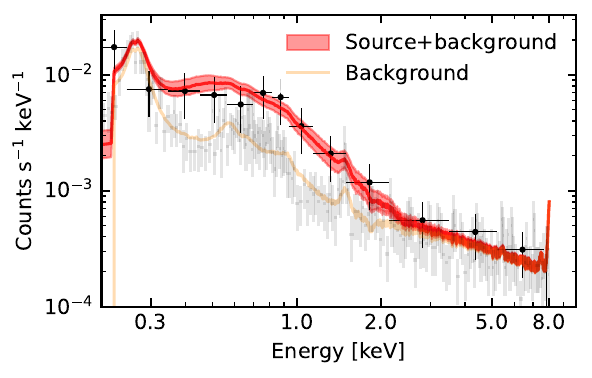}
    \caption{BXA fitted models to the convolved X-ray spectra from eROSITA (top, 14 days before optical peak) and \textit{XMM} (bottom two plots, $\sim$5~days and $\sim$180~days after optical peak). Black and grey markers represent source and scaled background spectra, respectively, with the background component not originating from the TDE host galaxy. The solid red line denotes the median model, whilst the shaded red band encloses 68\% of the posterior. The preferred model (Table~\ref{tab:x_ray_model_fits}) for the eRASS5 spectrum is thermal bremsstrahlung with $kT_{\mathrm{brems}}=71 ^{+8}_{-5}$~eV, whilst it is a powerlaw for the \textit{XMM} spectra, with $\Gamma$ of $2.7 ^{+0.3}_{-0.3}$ and $3.5 ^{+0.5}_{-0.5}$, respectively. The unconvolved models and spectra are presented in Fig.~\ref{fig:sed_evolution}.}
    \label{fig:erass5_fits}
\end{figure}

The eRASS5 spectrum, obtained $\sim$14 days before the optical peak, is ultra-soft, and can be well fitted by the thermal bremsstrahlung model with temperature $kT_{\mathrm{brems}}=71^{+8}_{-5}$~eV (Fig.~\ref{fig:erass5_fits}), or a blackbody with temperature $kT_{\mathrm{bb}}=47^{+5}_{-5}$~eV; such temperatures are consistent with the X-ray spectra of other X-ray bright thermal TDEs (e.g.~\citealt{saxton_x-ray_2020}). This corresponds to a 0.2--2~keV observed flux, $F_{\mathrm{X,\, obs}}=(3.4^{+0.6}_{-0.5})\times10^{-13}$~\unitflux, and a 0.2--2~keV unabsorbed flux, $F_{\mathrm{X,\, unabs}}=(1.6^{+0.4}_{-0.4})\times10^{-11}$~\unitflux ($L_{\mathrm{X}}=$\lumerassfive ~erg~s $^{-1}$). %TODO: add plot of the X-ray spectrum

\begin{figure*}
    \centering
    \includegraphics[scale=0.9]{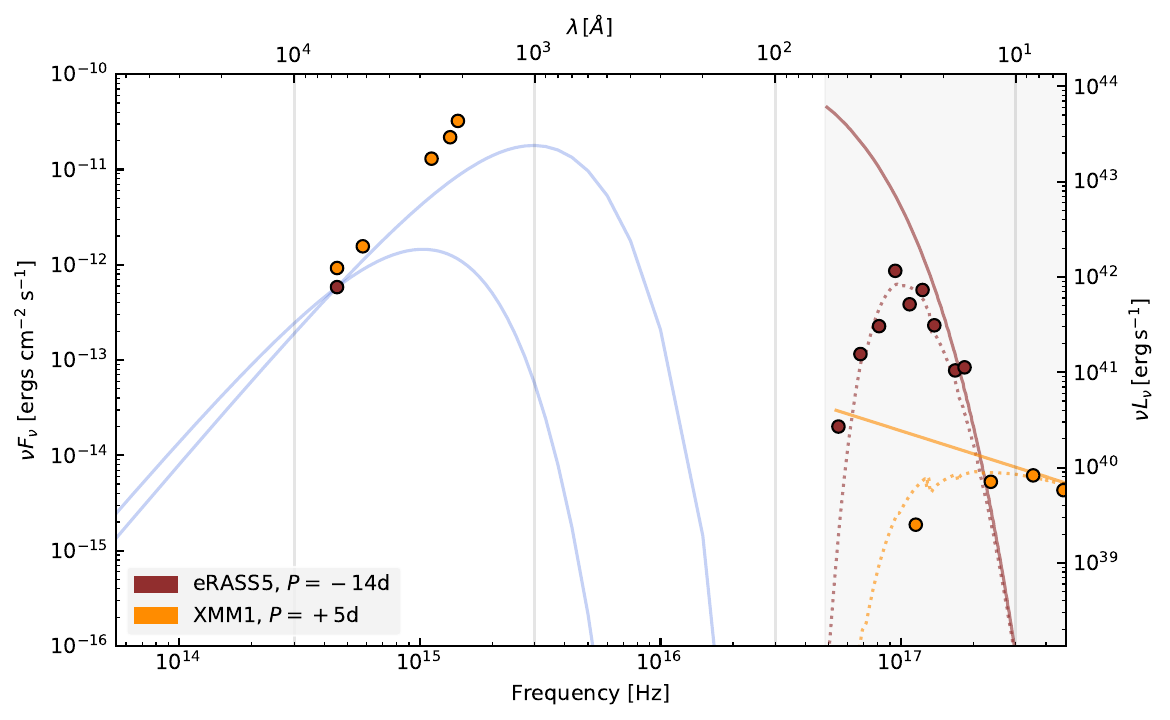}
    \caption{Early SED evolution of \dsb. The red and orange markers show the SED at the time of the eRASS5 detection (MJD$\sim$59627, 14 days before optical peak) and first \textit{XMM} observation $\sim$19 days later (MJD$\sim$59646, 5 days after optical peak). The dotted and solid lines in the X-ray band-pass (grey region) denote the observed and unabsorbed best fitting spectral models. The two blackbody curves (blue) passing through the ATLAS $o$-band data point on MJD$\sim$59627 are at temperatures of $\log [T /\mathrm{K} ]=4.1$ and $4.56$, the minimum and maximum temperatures inferred from fitting a single temperature blackbody to the multi-band photometry of a ZTF-selected TDE population \citep{hammerstein_final_2023}.} % log(T / Kelvin)=4.1, 4.56
    
    \label{fig:sed_evolution}
\end{figure*}

The first \textit{XMM} PN spectrum, taken $\sim$19 days after the eRASS5 spectrum, is harder, and can be best-fit by a power-law with photon index $2.7^{+0.3}_{-0.3}$. The eRASS5 to XMM spectral hardening is also accompanied by a factor of $\sim$\fxobsdrop\, decrease in $F_{\mathrm{X,\, obs}}$ to $(8.9^{+2.3}_{-1.8})\times 10^{-15}$~\unitflux. %The amplitude of the flux drop is even larger when considering $F_{\mathrm{X,\, unabs}}$, where it dims by a factor of $\sim$\fxunabsdrop\, to $F_{\mathrm{X,\, unabs}}=(3.5^{+0.9}_{-0.8})\times10^{-14}$~\unitflux over the 19 days between these two observations. 
The early-time evolution of the spectral energy distribution (SED) evolution between the eROSITA and XMM observation is plotted in Fig.~\ref{fig:sed_evolution}. 
At $\sim$173 ($\sim$154) days after the eRASS5 (first XMM observation), the second \textit{XMM} observation shows a power-law slope consistent with the first \textit{XMM} observation with photon index $3.5^{+0.5}_{-0.5}$, as well as a similar observed 0.2--2~keV flux of $(6.6^{+1.3}_{-1.1}) \times 10^{-15}$ \unitflux. The 0.2--2~keV fluxes in these \textit{XMM} observations are below the 3$\sigma$ upper limit inferred from the stacked eROSITA observations from its first four all-sky surveys (section~\ref{sec:discovery}). 
%{\color{red}At $\sim$173 ($\sim$154) days after the eRASS5 (first XMM observation), the second \textit{XMM} observation shows a slight re-softening of the X-ray emission, and is best fit by a power-law with photon index $4.7^{+0.9}_{-0.6}$. Although the observed 0.2--2~keV flux decreases slightly between the two \textit{XMM} observations to $F_{\mathrm{X,\, obs}}=6.7^{+1.3}_{-1.3}\times 10^{-15}$~\unitflux, $F_{\mathrm{X,\, unabs}}$ increases marginally by a factor of $\sim$3 to $1.0^{+0.9}_{-0.4}\times 10^{-13}$~\unitflux.}
%MJD & ObsID & \multicolumn{2}{c}{\texttt{zbbody}} & \multicolumn{2}{c}{\texttt{zpowerlaw}} & \multicolumn{2}{c}{\texttt{zbremsstrahlung}} \\
\begin{table*}
\centering
\caption{X-ray spectral fit results. The Akaike Information Criterion ($AIC$) column estimates the goodness of fit, with a lower $AIC$ representing a better fit.}%TODO: update script to auto add the correct obsids
\label{tab:x_ray_model_fits}
\begin{tabular}{r|cccccccccccccc}
\hline
MJD & ObsID & \multicolumn{2}{c}{\texttt{zbbody}} & \multicolumn{2}{c}{\texttt{zpowerlaw}} & \multicolumn{2}{c}{\texttt{zbremsstrahlung}} \\
\hline
 & & AIC & $kT$ [eV] & AIC & $\Gamma$ & AIC & $kT_{\mathrm{brems}}$ [eV] \\
\hline
59627.939 & eRASS5 & 367.6 & $47 ^{+5}_{-5}$ & 367.9 & $7.7 ^{+0.2}_{-0.4}$ & 366.8 & $71 ^{+8}_{-5}$ \\
59646.226 & XMM1 & 9908.3 & $219 ^{+28}_{-24}$ & 9893.8 & $2.7 ^{+0.3}_{-0.3}$ & 9899.7 & $768 ^{+346}_{-173}$ \\
59820.520 & XMM2 & 7925.8 & $150 ^{+26}_{-23}$ & 7918.0 & $3.5 ^{+0.5}_{-0.5}$ & 7921.1 & $403 ^{+155}_{-96}$ \\
\hline
\end{tabular}
\end{table*}

The photon index of $\Gamma \sim 2.7$ in the first \textit{XMM} observation is much softer than the photon indices of X-ray bright AGN; for example, \citet{nandra_ginga_1994} characterised a sample of continuum slopes of Seyfert galaxies by a Gaussian with $1.95 \pm 0.15$. In addition, it is also softer than the hard X-ray emission from an advection-dominated accretion flow (ADAF; e.g. \citealt{narayan_advection-dominated_1998}) of slope $\lesssim 2$ \citep{gu_anticorrelation_2009}, yet harder than the spectra of thermal TDEs \citep{saxton_correction_2021}. With a 0.2--2~keV luminosity of $\sim 4\times10^{40}$~erg~s$^{-1}$ and with no major change in flux in the 0.2--2 keV band between the two \textit{XMM} epochs, we consider the \textit{XMM} source spectra to be likely dominated by diffuse X-ray emission from within the circumnuclear environment of the host galaxy (i.e. unrelated to the TDE-triggered accretion episode onto the SMBH). This is in part motivated by the host galaxy of \dsb \, likely hosting a LLAGN prior to its 2022 outburst (recall the mid-infrared colour of $W1-W2\sim 0$ and previous type II AGN classification for its host galaxy; section~\ref{sec:discovery}), with past X-ray observations of nearby LLAGN also suggesting the presence of hot, diffuse plasma within a few hundred parsecs of the nucleus \citep{flohic_central_2006}, and with $\log [L_{0.5-2 \, \mathrm{keV}}]$ and $\log [L_{2-10  \, \mathrm{keV}}]$ luminosities in the range of $40.2 \pm 1.3 $ and $39.9 \pm 1.3$ \citep{gonzalez-martin_x-ray_2009}.
%This physical origin is preferred in Martin+ over an ULX/ XRB origin due to the mismatch of luminosity functions between the LLAGN and such systems.
%As the 0.2--2~keV fluxes in the \textit{XMM} observations are below the 3$\sigma$ upper limit inferred from the stacked eROSITA observations (section~\ref{sec:discovery}), the diffuse X-ray emission in the host galaxy was not previously detected by eROSITA.
%the scenario of a weak, constant, host galaxy hot gas emission being there all the time

A similar scenario may also have been present in the TDE candidate ASASSN-15oi \citep{gezari_x-ray_2017}, where two \textit{XMM} spectra, taken $\sim$80 days and $\sim$230 days after optical discovery, were best fitted\footnote{These spectral fit results are reported in \citealt{gezari_x-ray_2017} and were obtained using the first \textit{XMM} spectrum.} by a two component model consisting of a blackbody with $kT=47 \pm 3$~eV and a power-law with $\Gamma =2.5 \pm 0.8$. As ASASSN-15oi brightened in the X-rays over the $\sim 160$ days between these spectra, only the normalisation of the blackbody component increased, without a significant change in $kT$ or $\Gamma$. This would require a fair amount of model fine-tuning if the power-law component originated from Compton upscattered TDE disc photons, and the disc luminosity varied over time. Instead, this may be more easily explained if the disc emission evolves independently of the lower luminosity diffuse host emission (as also suggested in \citealt{gezari_x-ray_2017}), with the latter being emitted at much larger physical length scales than the X-ray emission from the TDE disc. 

\begin{figure}
    \centering
    \includegraphics[scale=0.8]{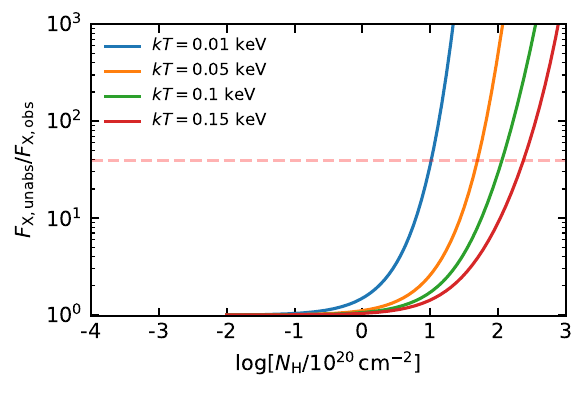}
    \includegraphics[scale=0.8]{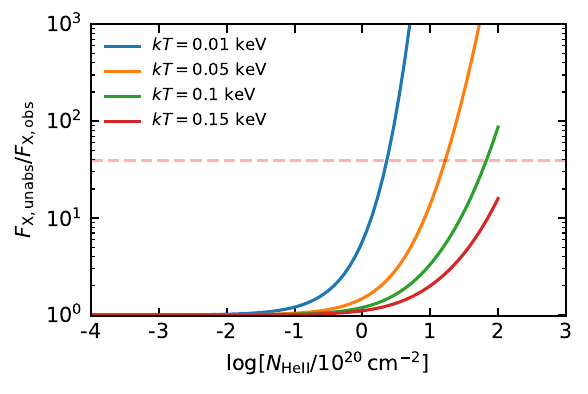}
    \caption{The amplitude of the observed flux drop in the 0.2--2~keV band ($F_{\mathrm{X, unabs}} / F_{\mathrm{X, obs}}$) due to absorption by neutral hydrogen (top) and \ion{He}{II} (bottom), with each curve representing a different blackbody temperature ($kT\sim 0.05 $~keV for AT~2022dsb). The sensitivity of $F_{\mathrm{X, unabs}} / F_{\mathrm{X, obs}}$ to absorption depending on the balance between the ionisation potential of the absorber and $kT$. The horizontal red dashed line corresponds to the flux drop by a factor of \fxobsdrop \, at early times seen in AT~2022dsb, corresponding to an $N_{\mathrm{H}}$ ($N_{\mathrm{HeII}}$) of $4.9 \times 10^{21}$ cm$^{-2}$ ($1.6 \times 10^{21}$ cm$^{-2}$).}
    \label{fig:absorber_xray_flux_drops}
\end{figure}

If the \textit{XMM} source spectra are dominated by the diffuse emission with the host galaxy, then the soft X-ray emission dominating the eRASS5 spectrum may have been obscured by optically thick material (see section~\ref{sec:discussion} for further discussion on its possible origin). Assuming that the TDE disc has spectral properties in its first \textit{XMM} spectrum similar to eRASS5 ($kT_{\mathrm{BB}} \sim 50$~eV) and a similar blackbody normalisation, then an increase in the neutral hydrogen column density along our line-of-sight to $> 4.9 \times 10^{21}$ cm$^{-2}$ would be capable of causing a 0.2—2 keV flux drop by a factor of at least \fxobsdrop \, (between these two spectra), or a \ion{He}{II} column density of $1.6 \times 10^{21}$ cm$^{-2}$ if the flux drop is due to photoionisation of \ion{He}{II} (Fig.~\ref{fig:absorber_xray_flux_drops}), when modelling \ion{He}{II} absorption\footnote{The column densities of all other species are set to zero within this absorption model here.} using the \texttt{xspec} model \texttt{ISMabs} \citep{gatuzz_ismabs_2015}. The conversion of these column densities into an estimate of the mass of a debris envelope obscuring the disc is complicated by the unknown ionisation fraction of helium in the debris; an alternate approach to constrain the reprocessor's properties at early times is presented in section~\ref{sec:discussion}.

\subsection{Photometric analysis}\label{sec:analysis_photometric}
\begin{figure}
    \centering
    \includegraphics[scale=0.8]{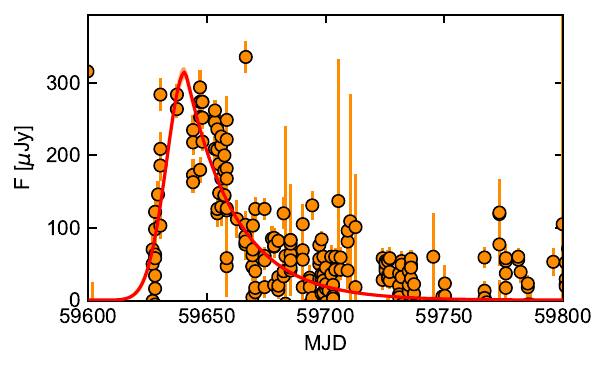}
    \caption{Half-Gaussian rise, exponential decay model (red) fitted to the ATLAS $o$-band photometry (orange markers). The shaded red bands denote the credible region enclosed by the 16$^{\mathrm{th}}$ and 84$^{\mathrm{th}}$ percentiles of the posterior. The datapoint with $F\sim300$~$\mu$Jy at MJD$\sim$59600 has a 1$\sigma$ error consistent with 0 (i.e. is not precursor emission to the main flare).}
    \label{fig:atlas_lightcurve_fitted}
\end{figure}

Following \citet{malyali_erasst_2023}, the ATLAS $o$-band lightcurves were fitted with a half-Gaussian rise, exponential decay model, as described in \citet{van_velzen_first_2019}, and plotted in Fig.~\ref{fig:atlas_lightcurve_fitted}. Only the $o$-band photometry was fitted here due to this providing the best sampling of the rise and decay of the optical emission, and only photometry in the range 59600$<$MJD$<$59750 was used. The inferred rise and decay timescales are $\sigma =7.9 ^{+0.4}_{-0.4}$~days and $\tau=21.0 ^{+0.6}_{-0.6}$~days, respectively, with the lightcurve peaking at MJD=\mjdpeak; this value is used as the reference time for the optical peak of \dsb\, in this work. The peak inferred $F_{\nu}$ is $320^{+7}_{-6}$~$\mu$Jy, corresponding to $\nu L _{\nu}\sim 2 \times 10^{42}$ erg~s$^{-1}$. To help understand when the eRASS5 detection occurs during the evolution of the TDE, then it is also valuable to define a start time for the optical rise based on the fitted model here. If one considers this to be when the optical flux is $\sim$1\% ($\sim 0.03$\%) of the optical peak, then this would occur when $\mathrm{MJD}_{\mathrm{start}}\sim$59617 ($\sim 59609$).
 % TODO: fix the y-axis label to flux density...

\begin{figure*}
    \centering
    \includegraphics[scale=0.8]{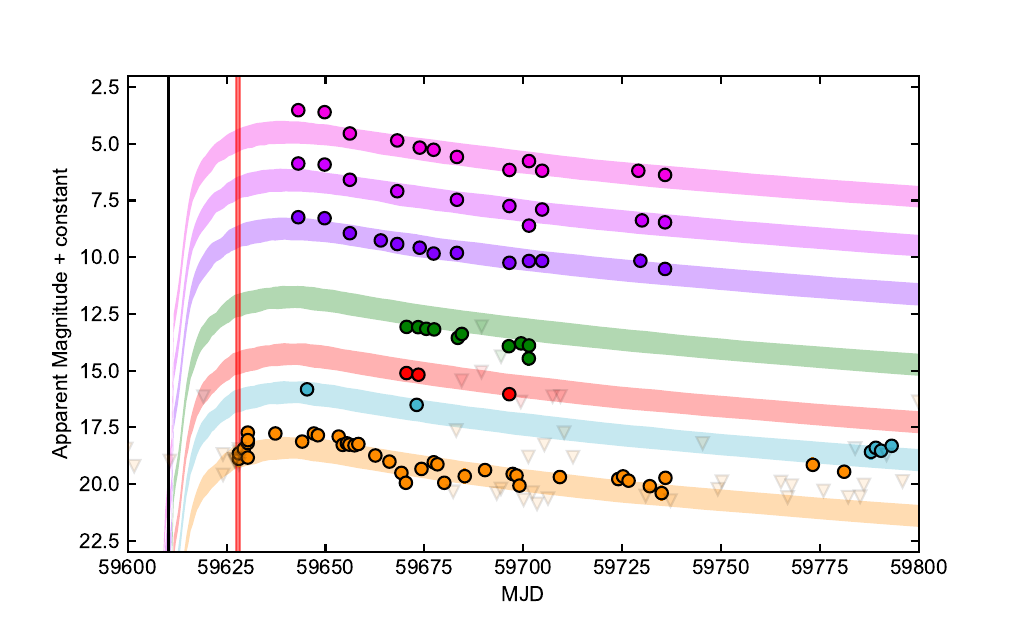}
    \caption{MOSFiT model fits to the multi-band photometry of \dsb , with colour scheme following Fig.~\ref{fig:multiwavelength_lc}. The black and red lines mark the estimated median time of first mass fallback ($\mathrm{MJD}_{\mathrm{fb}}=59610 \pm 6$) and the eRASS5 coverage of \dsb, occurring only $\sim$17 days later. A zoom-in on the ATLAS \textit{o}-band difference photometry sampling the optical rise, and which is used for constraining $\mathrm{MJD}_{\mathrm{fb}}$, is presented in Fig.~\ref{fig:atlas_lightcurve_fitted}.}
    \label{fig:mosfit_lightcurves}
\end{figure*}

To obtain a more physically-motivated estimate of the start time of the event, then we also fitted the multi-band photometry (ATLAS $o$ and $c$, ZTF $g$ and $r$, \textit{Swift} $UVW1$, $UVM2$ and $UVW2$ bands; Fig.~\ref{fig:mosfit_lightcurves}) with the TDE module \citep{mockler_weighing_2019} of the Modular OpenSource Fitter for Transients (MOSFiT; \citealt{guillochon_mosfit_2018}), using the nested sampler \texttt{dynesty} for posterior sampling \citep{speagle_dynesty_2020}. The free parameters of this model are the black hole mass ($M_{\mathrm{BH}}$), mass of the disrupted star ($M_{\star}$), the scaled impact parameter ($b$), the efficiency of converting accretion luminosity into the optical luminosity ($\epsilon$), a normalising factor and exponent for the photosphere radius ($R_{\mathrm{ph, 0}}$, $l_{\mathrm{ph}}$), a viscous delay timescale ($T_{\mathrm{viscous}}$) and the time of first massfall back to pericentre ($\mathrm{MJD}_{\mathrm{fb}}$). Although most of the inferred parameter estimates are dominated by systematics (Table~\ref{tab:mosfit_stats}), we note that the MOSFiT modelling does suggest a lower mass SMBH for the disruption with $\log (M_{\mathrm{bh}}/ M_{\odot})=(6.4 ^{+0.2}_{-0.2})\pm$0.2, as compared to the black hole mass derived from the $M_{\mathrm{BH}}-M_{\star}$ relation \citep{reines_relations_2015} using $M_{\star}$ from the host galaxy SED analysis (section~\ref{sec:discovery}). The inferred $\mathrm{MJD}_{\mathrm{fb}}=(59610.2 ^{+6.6}_{-6.3})\pm15$ is consistent with the start time for the optical rise inferred above. 
%\begin{itemize}
%    \item \textbf{Double-check the filters used by MOSFiT!!!}
%    \item The parameters inferred from this are listed in Table~\ref{tab:mosfit_stats}, and the fitted photometry is shown in Fig.~\ref{fig:mosfit_lightcurves}.
%    \item Of most interest is the $t_0$ parameter...
%    \item Is $t_{0, MOSFiT}$ consistent with the half-Gaussian model?
    %\item Early time X-ray rise prop to $t^2$? See 2020zso Wevers paper for references. 
    %\item Constant temperature rise (similar to 2019qiz- Nicholl+, Hung+)? Would allow us to measure the velocity of the outflow...
    %\item Lightcurve sub-structure around peak->see classifications of TDE lightcurves in literuature (Hammerstein+, Charalampopulous)?
%    \item \textbf{When did the source turn on? See discussion in perley cow papers} Time to rise from half max to max
%\end{itemize}
\begin{table}
\centering
\caption{Posterior medians and 1$\sigma$ credible regions inferred from the MOSFiT TDE lightcurve fitting. The estimated systematic errors on each estimate are taken from \citet{mockler_weighing_2019}.}
\label{tab:mosfit_stats}
\begin{tabular}{ccc}
\hline
Parameter & Value & Systematic Error\\
\hline
$\log (M_{\mathrm{bh}}/ M_{\odot})$ & $6.4 ^{+0.2}_{-0.2}$ & $\pm$0.2 \\
$\log (M_{\star}/ M_{\odot})$ & $0.2 ^{+0.3}_{-0.1}$ & $\pm$0.66 \\
$b$ & $1.0 ^{+0.2}_{-0.3}$ & $\pm$0.35 \\
$\log ( \epsilon )$ & $-2.0 ^{+0.5}_{-0.5}$ & $\pm$0.68 \\
$\log (R_{\mathrm{ph, 0}})$ & $-0.1 ^{+0.3}_{-0.3}$ & $\pm$0.4 \\
$l_{\mathrm{ph}}$ & $1.0 ^{+0.2}_{-0.2}$ & $\pm$0.2 \\
$\log (T_{\mathrm{viscous}} / \mathrm{days})$ & $-0.9 ^{+1.1}_{-1.2}$ & $\pm$0.1 \\
$\mathrm{MJD}_{\mathrm{fb}}$ & $59610.2 ^{+6.6}_{-6.3}$ & $\pm$15
\end{tabular}
\end{table}

\subsection{Spectroscopic analysis}
We used a modified version of the Python quasar fitting code (PyQSOFit; \citealt{guo_pyqsofit_2018}) to fit the optical spectra of \dsb \, after de-reddening the Galactic foreground contribution. First, we fitted the emission line free regions of each of the optical spectra with a power-law to estimate the continuum contribution to the spectrum, before dividing the observed spectrum by this continuum component to obtain a normalised spectrum. We note that we do not attempt to model or subtract the host galaxy component here, such that the modelled continuum involves both TDE and host galaxy emission. All spectral fitting made use of the python package \texttt{lmfit} \citep{newville_lmfit_2014} and the Markov Chain Monte Carlo (MCMC) sampler \texttt{emcee} \citep{foreman-mackey_emcee_2013}. 

\begin{figure}
    \centering
\includegraphics[scale=0.8]{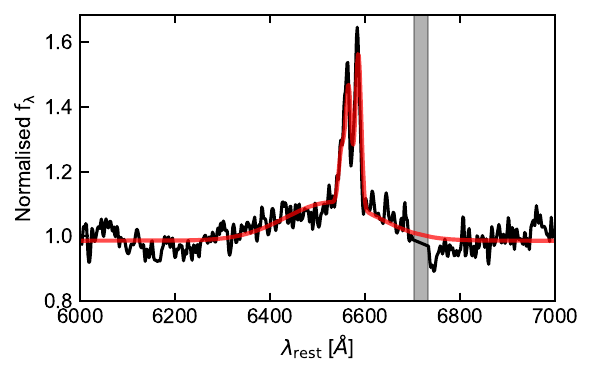}
    \caption{Example fit to the H$\alpha$ complex, with the continuum normalised spectrum in black and the best fitting model in red. The grey band denotes a region of telluric absorption which was masked during fitting. The centroid of the broad H$\alpha$ is shifted by \lcovelocityoffset ~km~s$^{-1}$ with respect to the rest frame of the host galaxy.}
    \label{fig:opspec_example_fit_halpha}
\end{figure}

After normalising by the continuum, we fitted each of the narrow emission lines in this complex (H$\alpha$, [\ion{N}{II}]~6548{\AA} and 6583~{\AA}) with a single Gaussian, and forced each of these to be of the same width. The broad H$\alpha$ component was fit with a single Gaussian (Fig~\ref{fig:opspec_example_fit_halpha}). Due to the possible presence and blending of emission from H$\beta$, \ion{He}{II}~4686{\AA}, \ion{N}{III} 4640{\AA}, H$\gamma$, \ion{Fe}{II} 37, 38, within the broad \ion{He}{II} complex, it is not straightforward to constrain the evolution of each of these possible components; we therefore examine the more isolated broad H$\alpha$ emission lines here. 

In the first LCO FLOYDS spectrum obtained $\sim$5 days before optical peak, the full-width half max (FWHM) of the broad H$\alpha$ line is \lcofwhm ~km~s$^{-1}$, and 
its centroid is blueshifted to $\lambda_{\mathrm{rest}}=$\lcocentroid ~{\AA}, corresponding to a velocity of \lcovelocityoffset ~km~s$^{-1}$. In the second optical spectrum obtained $\sim$1 day before optical peak, the FWHM is \nttfwhm ~km~s$^{-1}$, consistent with the earlier spectrum, but the velocity offset is \nttvelocityoffset ~km~s$^{-1}$. At later times, the H$\alpha$ emission line is not clearly seen above the host galaxy continuum emission. % TODO: flesh out this discussion more...

\section{early-time X-ray emission in TDEs}\label{sec:early_xray_emission_tdes}
In the following section, we briefly review the literature on early-time X-ray observations of TDEs, and compare the X-ray transient seen in AT~2022dsb (Fig.~\ref{fig:multiwavelength_lc}) with the wider TDE population.
%the wider-context of the discovery of the early-time X-ray transient in AT~2022dsb by briefly reviewing the 

The majority of the X-ray selected TDE population known prior to the launch of eROSITA was first discovered at a time when wide-field, high-cadence optical surveys were still relatively limited compared with the current generation, with respect to their depth, cadence, sky coverage, difference-imaging capabilities (important for nuclear transients), and ease-of-access to their optical lightcurves. Largely as a result of this, only a handful of the X-ray selected TDE candidates showed transient optical/ UV emission \citep{saxton_x-ray_2020}, which was only ever identified after the initial X-ray discovery, and typically only through \textit{Swift} UVOT follow-up. Furthermore, of the systems with detected transient optical/ UV emission, only the decaying phase of the TDE lightcurve was sampled in the UV (see discussion in Section 9 in a recent review \citealt{saxton_x-ray_2020}), thus the early-time X-ray evolution (during the initial optical/ UV brightening) of these X-ray selected systems remains unknown. Whilst the launch of eROSITA has seen a vast increase in the number of X-ray bright TDE candidates (e.g.~\citealt{malyali_at_2021,malyali_rebrightening_2023,malyali_erasst_2023,liu_deciphering_2023,homan_discovery_2023}), the majority of the TDE candidate population show no transient optical emission. In a sample of eROSITA-selected TDE candidates \citep{sazonov_first_2021}, only four systems display both transient X-ray and optical emission, but the detections of flaring X-ray emission associated to the TDE always occurs after the optical peak.  

\begin{figure*}
    \centering
    \includegraphics[scale=0.7]{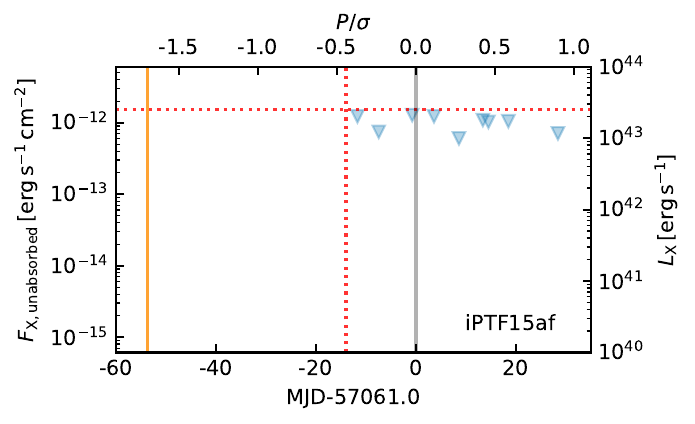}
    \includegraphics[scale=0.7]{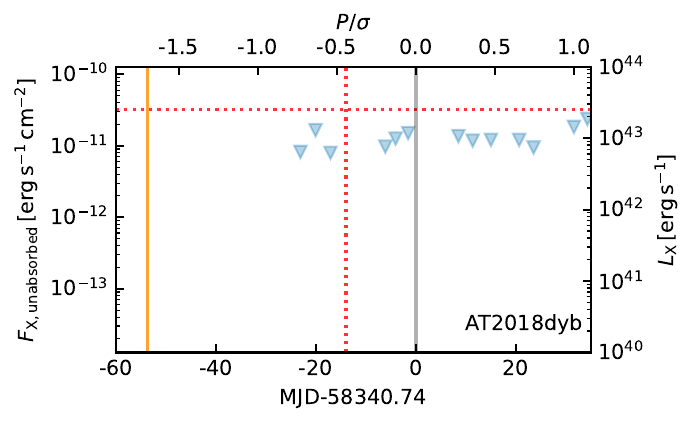}
    \includegraphics[scale=0.7]{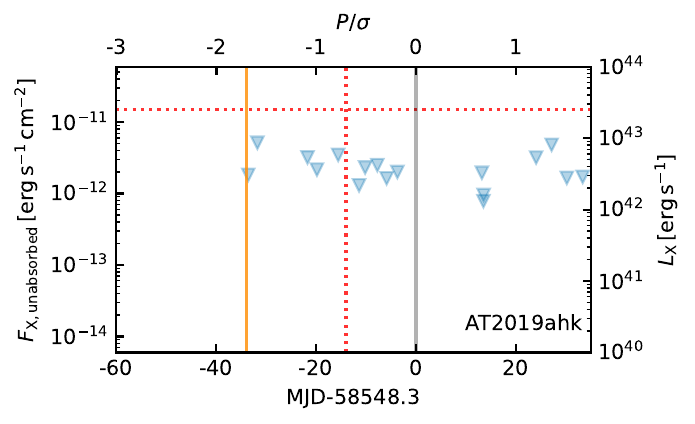}
    \includegraphics[scale=0.7]{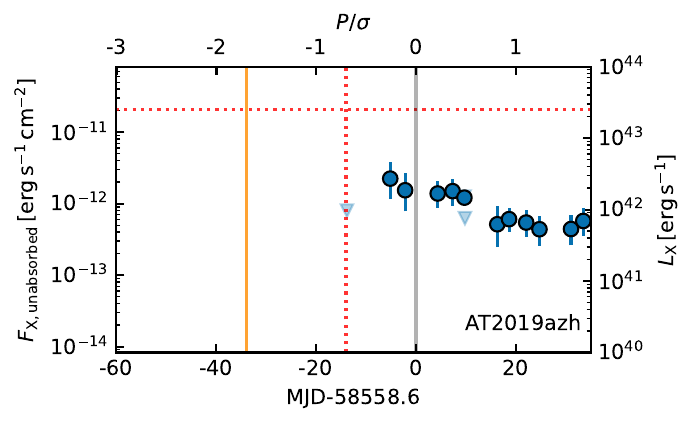}
    \includegraphics[scale=0.7]{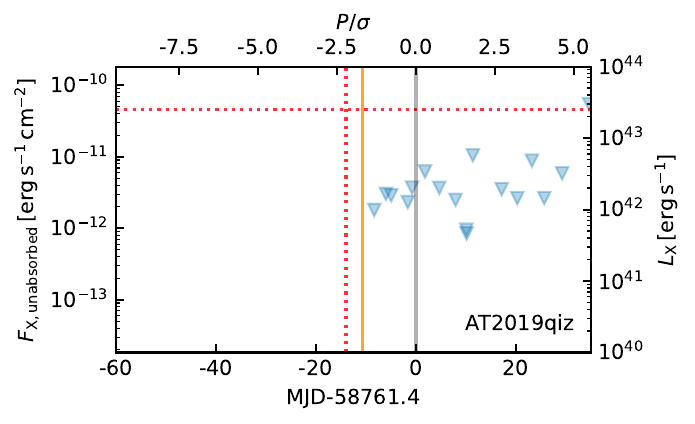}
    \includegraphics[scale=0.7]{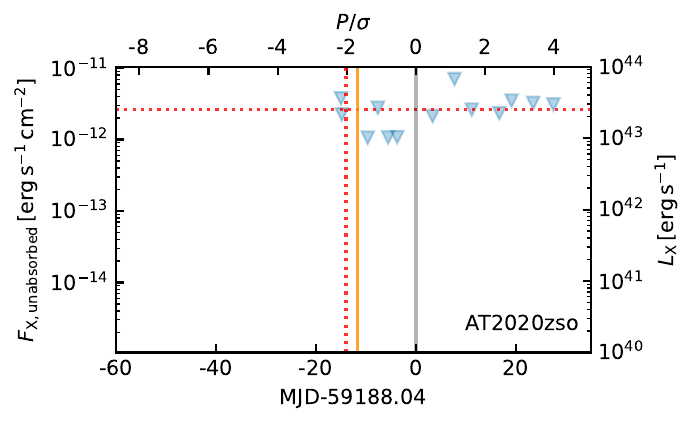}
    \caption{Early-time X-ray lightcurves of TDEs with observations pre-optical peak, with markers following the same definition as for Fig.~\ref{fig:multiwavelength_lc}. The dotted red lines mark the phase of the eRASS5 observation of \dsb \, and inferred 0.2--2~keV luminosity, whilst the dark orange solid line marks the eRASS5 observation at the time of its normed phase, $P/\sigma$ (Table~\ref{tab:early_time_xray_coverage}). MJDs are defined with respect to the inferred optical peaks in Table~\ref{tab:early_time_xray_coverage}. The eROSITA observation clearly represents the earliest X-ray detection of a TDE to date, and although there are XRT observations of these other TDEs at a comparable phase, only AT~2020zso has been observed at an earlier $P/\sigma$. 
    }
    \label{fig:early_xray_lightcurves}
\end{figure*}

Although the number of optically-selected TDE candidates has rapidly increased over the last decade, the majority of these have X-ray observations commencing at earliest near to, or after, peak optical brightness. This may in part stem from the very high discovery rate of transients in the latest generation of optical surveys, meaning that astronomers generally wait until close to peak optical brightness for an optical transient to become a strong TDE candidate, and only then trigger X-ray follow-up observations. Despite this, there are still \ntdeswithearlyxray\, optically-bright TDEs with X-ray observations starting before optical maximum (Table~\ref{tab:early_time_xray_coverage} and Fig.~\ref{fig:early_xray_lightcurves}), where this list was obtained via visual inspection of the joint X-ray and optical lightcurves of the TDEs presented in both the ZTF TDE sample \citep{hammerstein_final_2023}, and those in the recent TDE review paper by \citet{gezari_tidal_2021}. All of these TDEs are of H+He type\footnote{This is likely due current TDE follow-up strategies, since if a He complex is detected in a follow-up optical spectrum obtained during the optical rise, then there’s a stronger indication that the event is a TDE at these early times, and a higher likelihood of Swift XRT and UVOT observations being triggered with high urgency to monitor the evolution of the system. }, with the exception of AT~2019ahk, which only shows transient broad Balmer emission lines. For each of these systems, the 0.3--2~keV XRT lightcurves were generated and downloaded from the UKSSDC as in section~\ref{sec:obs_xray_xrt}. These were then converted to 0.2--2~keV lightcurves using webPIMMS, assuming a redshifted blackbody spectrum with $kT=50$~eV (similar to other X-ray bright TDEs; \citealt{saxton_x-ray_2020}), absorbed by a Galactic $N_{\mathrm{H}}$ along the line-of-sight to the TDE taken from \citet{willingale_calibration_2013}; lower $kT$ values for each TDE here would lead to higher estimated 0.2--2~keV fluxes.
% CAPTION: Properties of TDEs with X-ray observations pre-optical peak. MJD$_{\mathrm{peak}}$ is the inferred optical peak, $L^{\mathrm{1st}}_{\mathrm{X}}$ is the inferred 3$\sigma$ upper limit on the 0.2--2~keV luminosity from the earliest XRT observation and $P$ denotes the phase of the XRT observation relative MJD$_{\mathrm{peak}}$. $\sigma$ is the inferred rise timescale from a half-Gaussian fitted to the optical lightcurve (see section~\ref{sec:analysis_photometric}). MJD$_{\mathrm{peak}}$ and $\sigma$ were taken from \citet{van_velzen_optical-ultraviolet_2020} for iPTF15af, AT~2018dyb, AT~2019ahk, AT~2019azh, AT~2019qiz, and \citet{hammerstein_final_2022} for AT~2020zso.
\begin{table*}
\centering
\caption{Properties of TDEs with X-ray observations pre-optical peak. MJD$_{\mathrm{peak}}$ is the inferred optical peak, $L^{\mathrm{1st}}_{\mathrm{X}}$ is the inferred 3$\sigma$ upper limit on the 0.2--2~keV luminosity from the earliest X-ray observation (by \textit{Swift} XRT) and $P$ denotes the phase of the first X-ray observation relative to MJD$_{\mathrm{peak}}$. $\sigma$ is the inferred rise timescale from a half-Gaussian fitted to the optical lightcurve (section~\ref{sec:analysis_photometric}). MJD$_{\mathrm{peak}}$ and $\sigma$ were taken from \citet{van_velzen_optical-ultraviolet_2020} for iPTF15af, AT~2018dyb, AT~2019ahk, AT~2019azh, AT~2019qiz, and \citet{hammerstein_final_2023} for AT~2020zso.}
\label{tab:early_time_xray_coverage}
\begin{tabular}{cccccc}
\hline
Name & MJD$_{\mathrm{peak}}$ & $L^{\mathrm{1st}}_{\mathrm{X}}$ [erg~s$^{-1}$] & $P$ [d] & $\sigma$ [d] & $P / \sigma $ \\
\hline
AT2018dyb & $58340.7^{+1.4}_{-1.3}$ & $<7 \times 10^{42}$ & $-23.1$ & $31.6^{+2.3}_{-0.7}$ & $-0.7$ \\
AT2019ahk & $58548.3^{+0.8}_{-0.9}$ & $<3 \times 10^{42}$ & $-33.5$ & $20.0^{+0.5}_{-0.5}$ & $-1.7$ \\
iPTF15af & $57061.0^{+1.6}_{-1.8}$ & $<2 \times 10^{43}$ & $-11.7$ & $31.6^{+1.5}_{-2.1}$ & $-0.4$ \\
AT2019azh & $58558.6^{+1.5}_{-1.6}$ & $<1 \times 10^{42}$ & $-13.8$ & $20.0^{+2.4}_{-1.3}$ & $-0.7$ \\
AT2019qiz & $58761.4^{+0.6}_{-0.6}$ & $<1 \times 10^{42}$ & $-8.3$ & $6.3^{+0.1}_{-0.1}$ & $-1.3$ \\
AT2020zso & $59188.0^{+1.4}_{-1.4}$ & $<4 \times 10^{43}$ & $-15.0$ & $6.9^{+0.2}_{-0.2}$ & $-2.2$ \\
AT2022dsb & \mjdpeak & \lumerassfive & \phaseerassfivenoerr & $7.9 ^{+0.4}_{-0.4}$ & $-1.7$ \\
\end{tabular}
\end{table*}

\begin{table*}
\centering
\caption{Reported outflow properties of TDEs with X-ray observations pre-optical peak. Each of these objects have been followed up with different instruments and at different times, so there is not necessarily a common observed outflow indicator between systems. The expanding photosphere at early times points to a mechanical outflow only if it traces the motion of the debris (e.g. \citealt{nicholl_outflow_2020}).%\textbf{TODO: clarify photospheric expansion not necessairly mechanical outflow.
%Photosphere expansion implies a mechanical outflow at early times if the two expand together. At late times, the photosphere may start to contract whilst the the material outflow continues. 
%  From Wevers+22: "Alternatives to explain the photosphere expansion include the accumulation of matter around the peak of the mass fallback rate, which extends the photosphere to larger radii; it could be the result of time-dependent photon diffusion due to changing density and/or optical depth in the debris; or due to the orbital motion of heated matter, which at the inferred radius of ∼5 × 1014 cm is comparable (∼5000 km s−1) to the measured growth rate."
}
\label{tab:early_time_xray_outflow_properties}
\begin{tabular}{p{1.5cm}p{11cm}p{3cm}}
\hline
Name & Outflow properties & Reference \\
\hline
AT2018dyb & H$\alpha$ blueshifted by $\sim$700~km~s$^{-1}$ in spectrum obtained $\sim$24 days before peak (although H$\alpha$ redshifted after optical peak). & \citet{leloudas_spectral_2019} \\
AT2019ahk & Photosphere expanding at $\sim$2700~km~s$^{-1}$ via modelling photometric rise, assuming constant temperature. & \citet{holoien_discovery_2019}\\
iPTF15af & No H$\alpha$ detected in first optical spectrum near peak. \ion{Si}{IV} blueshifted by $\sim$6000~km~s$^{-1}$ in HST spectrum obtained $\sim$28~days post-peak. & \citet{blagorodnova_broad_2019} \\ % HST spectrum of 15af taken on 57089
AT2019azh & Transient radio emission detected $\sim$10 days before optical peak. & \citet{goodwin_at2019azh_2022}\\
AT2019qiz & Photosphere expanding at $\sim$2200~km~s$^{-1}$ via modelling photometric rise, assuming constant temperature. Inferred outflow velocity $\lesssim$5000~km~s$^{-1}$ from asymmetric H$\alpha$ line in spectrum $\sim$9 days before peak. & \citet{nicholl_outflow_2020} \\
AT2020zso & Photosphere expanding at $\sim$2900~km~s$^{-1}$ via modelling photometric rise, assuming constant temperature. % Radio emission detected on MJD=59204, but is this transient? https://www.wis-tns.org/astronotes/astronote/2021-24
& \citet{wevers_elliptical_2022} \\ 
\end{tabular}
\end{table*}

Comparing the X-ray lightcurve of \dsb\, with the other TDEs with pre-peak X-ray observations (Fig.~\ref{fig:early_xray_lightcurves}), then it is clear that the early-time transient X-ray emission in AT~2022dsb has never been observed before across all known TDEs with well sampled optical peaks. Of the TDEs in Table~\ref{tab:early_time_xray_coverage}, only AT~2019azh has a significant detection of soft X-ray emission before the observed optical maximum, with the system being detected for the first time $\sim$3~days before optical peak. AT~2019azh then remains at approximately a constant $L_{\mathrm{X}}$ over the following $\sim$40 days after the first significant detection (Fig.~\ref{fig:early_xray_lightcurves}), and thus shows a vastly different X-ray evolution to \dsb. 

The eRASS5 observation is not the earliest X-ray observation of a TDE in terms of the phase (number of days observed before optical maximum), with AT~2018dyb, AT~2019ahk, AT~2019azh and AT~2020zso all having been observed at earlier phases than \dsb. Furthermore, each of the earliest time observations for each system should also have been able to detect \dsb-like X-ray emission, given the 3$\sigma$ upper limits on $L_{\mathrm{X}}$ were lower than the $L_{\mathrm{X}}$ of the eRASS5 observation of \dsb. However, the optical-UV lightcurves of these TDEs evolve differently to \dsb, with respect to their peak luminosities and rise timescales. This is not unexpected, since these systems may span a range of different black hole masses, stellar masses and impact parameters. For example, the rise timescale of \dsb\, is inferred to be $\sim$8~days, whereas it is $\sim$31~days for iPTF-15af \citep{van_velzen_optical-ultraviolet_2020} (Table~\ref{tab:early_time_xray_coverage}). This complicates a clean comparison between these systems, and the task of understanding how early on in a TDE's evolution an \dsb-like X-ray transient may be observable. %As a result, this since a direct comparison between the lightcurves of TDEs is not trivial. 
If one considers the normalised phase (phase divided by the estimated rise timescale; Table~\ref{tab:early_time_xray_coverage}), then the eRASS5 observation of \dsb\, represents the second earliest X-ray observation of a TDE showing a transient \ion{He}{II} emission complex, with only AT~2020zso being observed at an earlier stage of the lightcurve.    

%\item AT 2020zso (MJD peak 59184.2 in Wevers+)
%\item However, the 3$\sigma$ upper limit on AT2020zso is 4.030438822866003e+43, which is above the eRASS5 $L_{\mathrm{X}}$ (i.e. the earliest XRT observation was not deep enough to detect the transient X-ray emission seen with eROSITA).
%\item The automated XRT lightcurve extraction pipeline actually splits up two observations performed on MJD=59173, with the second observation on this day deeper than the first. The inferred 3$\sigma$ upper limit on the count rate is $8\times10^{-3}$ counts s$^{-1}$, corresponding to $L_{\mathrm{X}}<2.4 \times 10^{43}$ erg/s, which is comparable to the eRASS5 $L_{\mathrm{X}}=2.3 \times 10^{43}$ erg/s. 
%\item A clean comparison of the normalised phase of AT~2022dsb with AT~2020zso is complicated by the optical-UV evolution during the latter's rise. 

Whilst the detection of the early-time X-ray transient in \dsb\, certainly benefitted from eROSITA serendipitously scanning over it a day before the first optical detection, this cannot be the sole factor in this discovery given the XRT coverage of other TDEs described above (i.e. there were observations that were early and deep enough to detect a source with spectral properties similar to \dsb , but did not because of physical differences between these systems). The fact that the X-ray emission could have been observed in other TDEs but was not, particularly for AT~2020zso (the TDE with observations performed at the earliest normed phase), suggests that the assumptions made when converting the observed 0.3--2~keV XRT count rate into an unabsorbed 0.2--2~keV flux (corrected only for Galactic absorption), and then a 0.2--2~keV intrinsic luminosity, may have been oversimplified and require further consideration.  
For example, an additional absorber along the line-of-sight to the TDE disc (such as from stellar debris ejected during the circularisation process, which may be optically thick or thin depending on the observer’s viewing angle to the system, or from neutral hydrogen in the host galaxy unrelated to the TDE), or a lower effective temperature of the disc emission (e.g. due to a retrograde black hole spin), would lead to larger estimated unabsorbed flux upper limits from the XRT observations, and might explain the previous non-detections of X-ray emission in these systems. The early-time X-ray transient seen in \dsb \, is likely not a universally observable feature in TDEs.

\section{Discussion}\label{sec:discussion}
The unique observational feature of \dsb \, with respect to the wider population of optically-selected TDEs is its early-time transient X-ray emission detected by eROSITA (see Table~\ref{tab:key_events} for a summary of the key events in the evolution of \dsb). From the physical modelling of the multi-band photometry (section~\ref{sec:analysis_photometric}), then the time of first mass fallback to pericentre after the disruption is estimated to be MJD$\sim 59610$ ($\sim$31 days before optical peak), meaning that the eROSITA discovery of ultra-soft X-ray emission on MJD 59627 occurs only 17 days after this. As the optical emission brightens in the system, then the observed X-ray emission in the 0.2--2~keV band decays over a 19 day period by a factor of
$\sim$\fxobsdrop \,(Fig.~\ref{fig:multiwavelength_lc}). This joint X-ray-to-optical evolution has not been observed before in a TDE candidate. Although the observed X-ray emission rapidly dims during the optical rise, \dsb\, shows a broad \ion{He}{II} complex which persists for at least $\sim$38 days after optical peak, and was first detected 5 days before peak (Fig.~\ref{fig:optical_spectroscopic_evolution}). Several outflow signatures are also present during the early stages of this TDE, in the form of blueshifted H$\alpha$ at \lcovelocityoffset ~km~s$^{-1}$, first observed at $P=-5$~days  (section~\ref{sec:spec_fitting}), radio transient emission at $P=20$~days (Fig.~\ref{fig:radiolc}), and blueshifted Ly$\alpha$ absorption lines at $\sim$\hstlymanalphavelocity ~km~s$^{-1}$, observed at $P=54$~days (\citealt{engelthaler_ultraviolet_2023}, Engelthaler et al., in prep.). Importantly, other past X-ray observations of He-TDEs before optical peak have not detected X-ray emission at a similar $L_{\mathrm{X}}$ (Table~\ref{tab:early_time_xray_coverage}), despite the observations being carried out at a similar phase to the eRASS5 observation of \dsb, and also having upper limits on $L_{\mathrm{X}}$ lower than for the eRASS5 detection of \dsb; each of the He TDEs in this sample have also been reported to show outflow signatures in observations performed around optical peak (Table~\ref{tab:early_time_xray_outflow_properties}). 
\begin{table}
    \centering
    %\begin{tabular}{c|l}
    \begin{tabular}{p{1cm}p{6.5cm}}
    \hline
       MJD  &  Event\\
       \hline
       $\sim$59610  & Time of first mass fallback to pericentre (section~\ref{sec:analysis_photometric}). \\
       59627 & eRASS5 detection. \\
       59628 & First 3$\sigma$ detection of optical emission in ATLAS $o$-band. \\
       59636 & Broad blueshifted H$\alpha$ and an emission complex detected around \ion{He}{II} 4686\AA, in the first follow-up optical spectrum of \dsb. \\
       59641 & Peak in the observed optical flux (section~\ref{sec:analysis_photometric}). \\
       59646 & First XMM observation, 0.2--2~keV observed flux drop by a factor of $\sim$\fxobsdrop \, relative to eRASS5. \\
        59661 & Detection of radio transient emission with ATCA in first radio follow-up observation. \\
        59693 & First detection of outflow at \hstlymanalphavelocity \, km~s$^{-1}$ from Ly$\alpha$ absorption (FWHM$\sim14000$~km~s$^{-1}$) in first HST spectrum \citep{engelthaler_ultraviolet_2023}.
    \end{tabular}
    \caption{Key events in the early evolution of \dsb.}
    \label{tab:key_events}
\end{table}

The early X-ray emission detected by eROSITA likely comes from an accretion disc that has recently been assembled through circularisation of the earliest-arriving gas in the fallback stream \citep{bonnerot_first_2021}. We rule out the early X-ray transient being produced by shock breakout emission from the surface of the star after being maximally compressed at pericentre \citep{carter_tidal_1983,guillochon_three-dimensional_2009,stone_consequences_2013,yalinewich_shock_2019}, as the predicted timescales of $\mathcal{O}$(minutes) for these flares are far shorter than what is observed in the eRASS5 observation ($>$1~day; Fig.~\ref{fig:xray_erass5_lightcurve}). %The shock breakout emission has also been predicted to be a hard X-ray source (REF), contrasting the early time soft emission observed in AT~2022dsb.
We also disfavour the X-ray transient being caused by accretion disc cooling (e.g.~\citealt{cannizzaro_accretion_2021}), since the disc temperature should increase as the optical emission brightens if the optical light curve traces the accretion rate at early times, or Lense-Thirring driven precession of the newly formed disc \citep{stone_observing_2012,franchini_lensethirring_2016}, as no rebrightening episodes are detected over the X-ray follow-up campaign (Fig.~\ref{fig:multiwavelength_lc}). 
Regardless of origin, then the high-energy tail of this early hard ionising source % 'this ?'
also likely drives the ionisation of \ion{He}{II} and the formation of the He-complex in the optical spectra.% (see section~\ref{sec:fluxdrop_alternatives} for alternate interpretations of the X-ray flux drop).

Around 14 days before the optical peak, it is likely that we have an unobscured view onto the nascent disc, which is initially surrounded by an envelope of gas formed through shocks during the circularization process, as found in simulations \citep{bonnerot_first_2021}. While this envelope is initially of low enough density for the disc emission to promptly emerge, %the arrival of more gas at later times increases the envelope mass. 
the envelope mass increases over time due to feeding by the outflowing gas in the system.
As a result, the X-ray emission may be more efficiently absorbed over time, leading to the observed drop-off in the X-ray emission, and a reprocessed-driven optical brightening. %The optical/ UV radiation produced from the reprocessing is then able to escape due to the lower optical depth to electron scattering. 
%To test this scenario, we computed the effective optical depth following \citet{roth_x-ray_2016}, finding that an increase to $\tau_{\rm eff} \approx 1$ (corresponding to efficient absorption) over the observed timescale of 19 days requires that the gas envelope gains mass at a rate $\dot{M}_{\rm env} \approx 5 M_\odot \rm yr^{-1}$, which is around the peak fallback rate for a typical TDE (e.g. \citealt{stone_stellar_2019}). % CB: I commented this passage to replace by an improved version
%{\color{red}TODO: add citation here, probably flesh out too, also mention metal absorption, lack of spherical symettry, probably add in plot sent through}
To quantitatively test this scenario, we follow \citet{roth_x-ray_2016} by modelling the envelope as a sphere of inner and outer radii $R_{\rm in}$ and $R_{\rm out}$, containing a mass of gas $M_{\rm env}$ distributed according to a density profile $\rho \propto R^{-2}$, which is irradiated by an inner source of luminosity $L$. The effective optical depth $\tau_{\rm eff}$ relevant for X-ray absorption is then given by their equation 27, relying on \ion{He}{II} photoionization being the dominant absorption process and using solar composition. Here, we further assume that the envelope mass increases with time as $M_{\rm env} = \dot{M}_{\rm env} t$ due to feeding by early returning debris. The time at which the envelope is able to absorb the inner X-ray radiation is obtained by solving $\tau_{\rm eff}(t) = 1$, which gives 
\begin{multline}
t_{\rm abs} = 24 \, {\rm d}  \, \left(\frac{L}{10^{43} \, \rm{erg \, s^{-1}}} \right)^{5/19} \left(\frac{\dot{M}_{\rm env}}{M_{\odot} \, \rm{ yr^{-1}}}\right)^{-1} \\ \left(\frac{R_{\rm in}}{10^{14} \, \rm{cm}} \right)^{20/19} \left(\frac{R_{\rm out}}{10^{15} \, \rm{cm}} \right)^{16/19},
\end{multline}
where the luminosity $L \approx 10^{43} \, \rm{erg \, s^{-1}}$ is determined from that of the detected X-rays. This time is approximately consistent with the 19 day delay between the eRASS5 detection and first \textit{XMM} observation, for an envelope feeding rate $\dot{M}_{\rm env} \approx  M_{\odot} \, \rm{ yr^{-1}}$, comparable to the debris fallback rate of a typical TDE near peak (e.g. \citealt{rossi_process_2021}). Within this interpretation, the early-time X-ray detection presented in this paper provides a new way to constrain physical properties such as the feeding rate and size of the envelope, and the luminosity of the obscured accretion disc, which are crucial to improve our theoretical understanding of these systems.

The origin of the early-time outflows seen in some TDEs is still unclear (see discussion in \citealt{goodwin_at2019azh_2022}), but is thought to be due to either a stream-stream collision-induced outflow (CIO; e.g.~\citealt{lu_self-intersection_2020}), debris unbounded by accretion luminosity \citep{metzger_bright_2016}  or a radiatively-driven disc wind \citep{lodato_multiband_2011,miller_disk_2015}. The data set in this work does not allow us to distinguish between these mechanisms for \dsb, since each scenario would initially lead to an increased density of gas and optical depth along our line-of-sight as the fallback rate increases over time\footnote{Future spectroscopic monitoring of TDEs in the UV may distinguish between these two origins.}. 
%Outflow: likely non-spherically symmetric to explain why X-ray emission from 2022dsb was seen but not in other TDEs with comparable X-ray observations (section~\ref{sec:early_xray_emission_tdes}), when it should have been. 

Importantly, if the observed early-time evolution in \dsb\, is not unique to this system, then other TDEs may also be X-ray bright at early times, and become X-ray faint only when veiled by outflowing debris launched shortly after the onset of circularisation.\footnote{Alternatively, if the X-ray emission is above a critical luminosity, then all of the \ion{He}{II} may be ionised to \ion{He}{III}, enabling the X-ray emission to escape the system \citep{metzger_bright_2016}.}. Given that the existing models described above predict the launching of outflows which would extend large solid angles on the sky, as seen by the disrupting black hole, then a large fraction of optically-bright TDEs may therefore be X-ray faint when followed up in the weeks-to-months after optical peak (unless viewed at angles peering through an optically thin funnel in the reprocessor; \citealt{metzger_bright_2016,dai_unified_2018,lu_self-intersection_2020}).

\section{Summary}\label{sec:conclusions}
We reported on multi-wavelength observations of the TDE candidate \dsb, whose main properties can be summarised as follows:
\begin{enumerate}
    \item eROSITA detected ultra-soft ($kT_{\mathrm{BB}} \sim 45$~eV) X-ray emission (0.2--2~keV $L_{\mathrm{X}}\sim 3 \times10^{43}$ erg~s$^{-1}$)
    from a TDE $\sim$14 days before optical peak. The eROSITA detection precedes the first 3$\sigma$ detection in the optical, and occurs only $\sim$17 days after the inferred time of first mass fallback to pericentre.
    \item An \textit{XMM} follow-up observation 19 days after this eROSITA detection revealed a drop in the observed 0.2--2~keV flux by a factor of \fxobsdrop ; during this period, the optical emission brightened to a maximum. A second \textit{XMM} observation $\sim$173 days after the eROSITA detection showed a 0.2--2~keV flux and spectral properties consistent with this first \textit{XMM} observation. No further X-ray emission was significantly detected above background by \textit{Swift} XRT monitoring observations in the following $\sim$200 days after the eROSITA detection.  Thus without the early-time eROSITA observation, AT~2022dsb would likely have been classified as an `optically bright, X-ray quiet' TDE.
    \item Follow-up optical spectra show a broad emission complex around the \ion{He}{II}~4686{\AA}, broad H$\alpha$ emission and a strong blue continuum in the early-time spectra. The \ion{He}{II} complex is clearly present in the spectra taken $\sim$5 days before optical peak, and is still detected $\sim$38 days after optical peak (even after the large amplitude X-ray dimming). The strength of these features with respect to the host galaxy emission decreases over the spectroscopic follow-up campaign.
    \item Multiple outflow signatures are detected in the system at early times (transient radio emission with ATCA, first detected $\sim$20 days post-optical peak; blueshifted broad H$\alpha$ emission at $\sim$1600~km~s$^{-1}$, detected $\sim$5~days before optical peak, and blueshifted broad Ly$\alpha$ absorption at $\sim-$\hstlymanalphavelocity \,km~s$^{-1}$, detected $\sim$54 days after optical peak). 
    \item The combination of these observed features suggests that outflows launched at early times %As the mass fallback rate rises to a maximum at early times post-disruption, then the early massive outflows (linked to the increasing mass fallback rate) 
    may boost the density of the material enshrouding the nascent disc, leading to an increased amount of reprocessing of the high-energy disc emission. This causes an early drop-off in the observed X-ray flux whilst the optical brightens. 
    \item If the observed early-time properties are not unique to this system, then other TDEs may be X-ray bright at early times, and become X-ray faint when veiled by outflowing stellar debris. The X-ray vs optically bright nature of a TDE is also time dependent at early times.
\end{enumerate}

The early-time X-ray emission from TDEs may be monitored in greater detail with the next-generation of time-domain missions, such as the \textit{Einstein Probe} (\textit{EP}; \citealt{yuan_einstein_2018}), scheduled for launch in late 2023, or through early follow-up of candidates identified with the \textit{Ultraviolet Transient Astronomy Satellite} (\textit{ULTRASAT}; \citealt{sagiv_science_2014}) and the Vera Rubin Observatory \citep{ivezic_lsst_2019}. High cadence X-ray monitoring observations of such early X-ray transients may provide a new way to constrain the mass feeding rate and nature of the reprocessing envelope in TDEs in future work. %%Within this interpretation, the early-time X-ray detection presented in this paper provides a new way to constrain physical properties such as the feeding rate and size of the envelope, and the luminosity of the obscured accretion disc, which are crucial to improve our theoretical understanding of these systems. 

\section*{Acknowledgements}
AM is grateful to the generosity of Curtin University for hosting his visit, where parts of this work were completed. AM thanks the \textit{XMM}, \textit{Swift} and \textit{NICER} teams for approving the ToO requests.  AM acknowledges support by DLR under the grant 50 QR 2110 (XMM\_NuTra, PI: Z. Liu). This work was performed in part at the Aspen Center for Physics, which is supported by National Science Foundation grant PHY-2210452.

This work is based on data from eROSITA, the soft X-ray instrument aboard SRG, a joint Russian-German science mission supported by the Russian Space Agency (Roskosmos), in the interests of the Russian Academy of Sciences represented by its Space Research Institute (IKI), and the Deutsches Zentrum für Luft- und Raumfahrt (DLR). The SRG spacecraft was built by Lavochkin Association (NPOL) and its subcontractors, and is operated by NPOL with support from the Max Planck Institute for Extraterrestrial Physics (MPE).

The development and construction of the eROSITA X-ray instrument was led by MPE, with contributions from the Dr. Karl Remeis Observatory Bamberg \& ECAP (FAU Erlangen-Nuernberg), the University of Hamburg Observatory, the Leibniz Institute for Astrophysics Potsdam (AIP), and the Institute for Astronomy and Astrophysics of the University of T{\"u}bingen, with the support of DLR and the Max Planck Society. The Argelander Institute for Astronomy of the University of Bonn and the Ludwig Maximilians Universit{\"a}t Munich also participated in the science preparation for eROSITA.

The eROSITA data shown here were processed using the eSASS software system developed by the German eROSITA consortium.

The authors acknowledge support for obtaining the LCO/FLOYDS spectroscopy by the Deutsche Forschungsgemeinschaft (DFG, German Research Foundation) under Germany’s Excellence Strategy - EXC-2094 - 390783311.

This work made use of data supplied by the UK Swift Science Data Centre at the University of Leicester.

%%%%ATCA acknowledgement

The Australia Telescope Compact Array is part of the Australia Telescope National Facility\footnote{\url{https://ror.org/05qajvd42}} which is funded by the Australian Government for operation as a National Facility managed by CSIRO. We acknowledge the Gomeroi people as the Traditional Owners of the Observatory site.

%%%Start of legacy acknowledgement
The Legacy Surveys consist of three individual and complementary projects: the Dark Energy Camera Legacy Survey (DECaLS; Proposal ID 2014B-0404; PIs: David Schlegel and Arjun Dey), the Beijing-Arizona Sky Survey (BASS; NOAO Prop. ID 2015A-0801; PIs: Zhou Xu and Xiaohui Fan), and the Mayall z-band Legacy Survey (MzLS; Prop. ID 2016A-0453; PI: Arjun Dey). DECaLS, BASS and MzLS together include data obtained, respectively, at the Blanco telescope, Cerro Tololo Inter-American Observatory, NSF’s NOIRLab; the Bok telescope, Steward Observatory, University of Arizona; and the Mayall telescope, Kitt Peak National Observatory, NOIRLab. Pipeline processing and analyses of the data were supported by NOIRLab and the Lawrence Berkeley National Laboratory (LBNL). The Legacy Surveys project is honored to be permitted to conduct astronomical research on Iolkam Du’ag (Kitt Peak), a mountain with particular significance to the Tohono O’odham Nation.

Some of the observations reported in this paper were obtained with the Southern African Large Telescope (SALT) under the programme 2021-2-LSP-001 (PI: DAHB). Polish participation in SALT is funded by grant No.\ MEiN nr2021/WK/01.

NOIRLab is operated by the Association of Universities for Research in Astronomy (AURA) under a cooperative agreement with the National Science Foundation. LBNL is managed by the Regents of the University of California under contract to the U.S. Department of Energy.

This project used data obtained with the Dark Energy Camera (DECam), which was constructed by the Dark Energy Survey (DES) collaboration. Funding for the DES Projects has been provided by the U.S. Department of Energy, the U.S. National Science Foundation, the Ministry of Science and Education of Spain, the Science and Technology Facilities Council of the United Kingdom, the Higher Education Funding Council for England, the National Center for Supercomputing Applications at the University of Illinois at Urbana-Champaign, the Kavli Institute of Cosmological Physics at the University of Chicago, Center for Cosmology and Astro-Particle Physics at the Ohio State University, the Mitchell Institute for Fundamental Physics and Astronomy at Texas A\&M University, Financiadora de Estudos e Projetos, Fundacao Carlos Chagas Filho de Amparo, Financiadora de Estudos e Projetos, Fundacao Carlos Chagas Filho de Amparo a Pesquisa do Estado do Rio de Janeiro, Conselho Nacional de Desenvolvimento Cientifico e Tecnologico and the Ministerio da Ciencia, Tecnologia e Inovacao, the Deutsche Forschungsgemeinschaft and the Collaborating Institutions in the Dark Energy Survey. The Collaborating Institutions are Argonne National Laboratory, the University of California at Santa Cruz, the University of Cambridge, Centro de Investigaciones Energeticas, Medioambientales y Tecnologicas-Madrid, the University of Chicago, University College London, the DES-Brazil Consortium, the University of Edinburgh, the Eidgenossische Technische Hochschule (ETH) Zurich, Fermi National Accelerator Laboratory, the University of Illinois at Urbana-Champaign, the Institut de Ciencies de l’Espai (IEEC/CSIC), the Institut de Fisica d’Altes Energies, Lawrence Berkeley National Laboratory, the Ludwig Maximilians Universitat Munchen and the associated Excellence Cluster Universe, the University of Michigan, NSF’s NOIRLab, the University of Nottingham, the Ohio State University, the University of Pennsylvania, the University of Portsmouth, SLAC National Accelerator Laboratory, Stanford University, the University of Sussex, and Texas A\&M University.

BASS is a key project of the Telescope Access Program (TAP), which has been funded by the National Astronomical Observatories of China, the Chinese Academy of Sciences (the Strategic Priority Research Program “The Emergence of Cosmological Structures” Grant XDB09000000), and the Special Fund for Astronomy from the Ministry of Finance. The BASS is also supported by the External Cooperation Program of Chinese Academy of Sciences (Grant 114A11KYSB20160057), and Chinese National Natural Science Foundation (Grant 12120101003, 11433005).

The Legacy Survey team makes use of data products from the Near-Earth Object Wide-field Infrared Survey Explorer (NEOWISE), which is a project of the Jet Propulsion Laboratory/California Institute of Technology. NEOWISE is funded by the National Aeronautics and Space Administration.

The Legacy Surveys imaging of the DESI footprint is supported by the Director, Office of Science, Office of High Energy Physics of the U.S. Department of Energy under Contract No. DE-AC02-05CH1123, by the National Energy Research Scientific Computing Center, a DOE Office of Science User Facility under the same contract; and by the U.S. National Science Foundation, Division of Astronomical Sciences under Contract No. AST-0950945 to NOAO.%%%End of legacy acknowledgement

This work has made use of data from the Asteroid Terrestrial-impact Last Alert System (ATLAS) project. The Asteroid Terrestrial-impact Last Alert System (ATLAS) project is primarily funded to search for near earth asteroids through NASA grants NN12AR55G, 80NSSC18K0284, and 80NSSC18K1575; byproducts of the NEO search include images and catalogs from the survey area. This work was partially funded by Kepler/K2 grant J1944/80NSSC19K0112 and HST GO-15889, and STFC grants ST/T000198/1 and ST/S006109/1. The ATLAS science products have been made possible through the contributions of the University of Hawaii Institute for Astronomy, the Queen’s University Belfast, the Space Telescope Science Institute, the South African Astronomical Observatory,  and The Millennium Institute of Astrophysics (MAS), Chile.

This work was supported by the Australian government through the Australian Research Council's Discovery Projects funding scheme (DP200102471).

%%%%%%%%%%%%%%%%%%%%%%%%%%%%%%%%%%%%%%%%%%%%%%%%%%
\section*{Data Availability}
A public release of the entire eRASS1 data taken within the German half of the eROSITA sky is anticipated for Q4 2023. eRASS2-5
data are expected to be released at a later stage.
%The eRASS1-4 data taken within the German half of the eROSITA sky is currently planned to be made public by Q2 2024, whilst the eRASS5 data is scheduled to become public by Q2 2026. 
The Swift data is available to download through the UK Swift Data Science website\footnote{\url{https://www.swift.ac.uk/archive/index.php}}. The \textit{XMM} data will become public after the propietory period expires (2023-09-21). Publicly available ATLAS data can be accessed through the ATLAS forced photometry service\footnote{\url{https://fallingstar-data.com/forcedphot/}}. Publicly available ZTF data can be accessed through the ZTF forced photometry service\footnote{\url{https://irsa.ipac.caltech.edu/Missions/ztf.html}}.
NTT/EFOSC2 spectroscopy has been obtained under the program IDs 108.220C.012 (PI. C. Inserra) and 109.23JL.001 (PI. I. Grotova). All optical spectra are publicly available.
%Follow-up optical spectra will likely remain private at least until the release of the forthcoming eROSITA-selected TDE population paper, but could be made available upon reasonable request. 
ATCA data are stored in the Australia Telescope Online Archive\footnote{\url{https://atoa.atnf.csiro.au/}}, and will become publicly accessible 18 months from the date of observation.



\bibliographystyle{mnras}
\bibliography{at2022dsb} % if your bibtex file is called example.bib

\begin{thebibliography}{}
\makeatletter
\relax
\def\mn@urlcharsother{\let\do\@makeother \do\$\do\&\do\#\do\^\do\_\do\%\do\~}
\def\mn@doi{\begingroup\mn@urlcharsother \@ifnextchar [ {\mn@doi@}
  {\mn@doi@[]}}
\def\mn@doi@[#1]#2{\def\@tempa{#1}\ifx\@tempa\@empty \href
  {http://dx.doi.org/#2} {doi:#2}\else \href {http://dx.doi.org/#2} {#1}\fi
  \endgroup}
\def\mn@eprint#1#2{\mn@eprint@#1:#2::\@nil}
\def\mn@eprint@arXiv#1{\href {http://arxiv.org/abs/#1} {{\tt arXiv:#1}}}
\def\mn@eprint@dblp#1{\href {http://dblp.uni-trier.de/rec/bibtex/#1.xml}
  {dblp:#1}}
\def\mn@eprint@#1:#2:#3:#4\@nil{\def\@tempa {#1}\def\@tempb {#2}\def\@tempc
  {#3}\ifx \@tempc \@empty \let \@tempc \@tempb \let \@tempb \@tempa \fi \ifx
  \@tempb \@empty \def\@tempb {arXiv}\fi \@ifundefined
  {mn@eprint@\@tempb}{\@tempb:\@tempc}{\expandafter \expandafter \csname
  mn@eprint@\@tempb\endcsname \expandafter{\@tempc}}}

\bibitem[\protect\citeauthoryear{Alexander, Berger, Guillochon, Zauderer  \&
  Williams}{Alexander et~al.}{2016}]{alexander_discovery_2016}
Alexander K.~D.,  Berger E.,  Guillochon J.,  Zauderer B.~A.,   Williams P.
  K.~G.,  2016, \mn@doi [The Astrophysical Journal]
  {10.3847/2041-8205/819/2/L25}, 819, L25

\bibitem[\protect\citeauthoryear{Alexander, Wieringa, Berger, Saxton  \&
  Komossa}{Alexander et~al.}{2017}]{alexander_radio_2017}
Alexander K.~D.,  Wieringa M.~H.,  Berger E.,  Saxton R.~D.,   Komossa S.,
  2017, \mn@doi [The Astrophysical Journal] {10.3847/1538-4357/aa6192}, 837,
  153

\bibitem[\protect\citeauthoryear{Anderson et~al.,}{Anderson
  et~al.}{2020}]{anderson_caltech-nrao_2020}
Anderson M.~M.,  et~al., 2020, \mn@doi [The Astrophysical Journal]
  {10.3847/1538-4357/abb94b}, 903, 116

\bibitem[\protect\citeauthoryear{Arcavi et~al.,}{Arcavi
  et~al.}{2014}]{arcavi_continuum_2014}
Arcavi I.,  et~al., 2014, \mn@doi [The Astrophysical Journal]
  {10.1088/0004-637X/793/1/38}, 793, 38

\bibitem[\protect\citeauthoryear{Assef, Stern, Noirot, Jun, Cutri  \&
  Eisenhardt}{Assef et~al.}{2018}]{assef_wise_2018}
Assef R.~J.,  Stern D.,  Noirot G.,  Jun H.~D.,  Cutri R.~M.,   Eisenhardt P.
  R.~M.,  2018, \mn@doi [The Astrophysical Journal Supplement Series]
  {10.3847/1538-4365/aaa00a}, 234, 23

\bibitem[\protect\citeauthoryear{Bade, Komossa  \& Dahlem}{Bade
  et~al.}{1996}]{bade_detection_1996}
Bade N.,  Komossa S.,   Dahlem M.,  1996, Astronomy \& Astrophysics, 309, L35

\bibitem[\protect\citeauthoryear{Bellm et~al.,}{Bellm
  et~al.}{2019}]{bellm_zwicky_2019}
Bellm E.~C.,  et~al., 2019, \mn@doi [Publications of the Astronomical Society
  of the Pacific] {10.1088/1538-3873/aaecbe}, 131, 018002

\bibitem[\protect\citeauthoryear{Blagorodnova et~al.,}{Blagorodnova
  et~al.}{2017}]{blagorodnova_iptf16fnl_2017}
Blagorodnova N.,  et~al., 2017, \mn@doi [The Astrophysical Journal]
  {10.3847/1538-4357/aa7579}, 844, 46

\bibitem[\protect\citeauthoryear{Blagorodnova et~al.,}{Blagorodnova
  et~al.}{2019}]{blagorodnova_broad_2019}
Blagorodnova N.,  et~al., 2019, \mn@doi [The Astrophysical Journal]
  {10.3847/1538-4357/ab04b0}, 873, 92

\bibitem[\protect\citeauthoryear{Bonnerot, Lu  \& Hopkins}{Bonnerot
  et~al.}{2021}]{bonnerot_first_2021}
Bonnerot C.,  Lu W.,   Hopkins P.~F.,  2021, \mn@doi [Monthly Notices of the
  Royal Astronomical Society] {10.1093/mnras/stab398}, 504, 4885

\bibitem[\protect\citeauthoryear{Brown et~al.,}{Brown
  et~al.}{2013}]{brown_cumbres_2013}
Brown T.~M.,  et~al., 2013, \mn@doi [Publications of the Astronomical Society
  of the Pacific] {10.1086/673168}, 125, 1031

\bibitem[\protect\citeauthoryear{Brunner et~al.,}{Brunner
  et~al.}{2022}]{brunner_erosita_2022}
Brunner H.,  et~al., 2022, \mn@doi [Astronomy \& Astrophysics]
  {10.1051/0004-6361/202141266}, 661, A1

\bibitem[\protect\citeauthoryear{Buchner}{Buchner}{2021}]{buchner_ultranest_2021}
Buchner J.,  2021, {UltraNest} -- a robust, general purpose {Bayesian}
  inference engine, \url {http://arxiv.org/abs/2101.09604}

\bibitem[\protect\citeauthoryear{Buchner et~al.,}{Buchner
  et~al.}{2014}]{buchner_x-ray_2014}
Buchner J.,  et~al., 2014, \mn@doi [Astronomy \& Astrophysics]
  {10.1051/0004-6361/201322971}, 564, A125

\bibitem[\protect\citeauthoryear{Buckley, Swart  \& Meiring}{Buckley
  et~al.}{2006}]{buckley_completion_2006}
Buckley D. A.~H.,  Swart G.~P.,   Meiring J.~G.,  2006, in Stepp L.~M.,  ed.,
  Society of {Photo}-{Optical} {Instrumentation} {Engineers} ({SPIE})
  {Conference} {Series} Vol. 6267, Society of {Photo}-{Optical}
  {Instrumentation} {Engineers} ({SPIE}) {Conference} {Series}. p. 62670Z,
  \mn@doi{10.1117/12.673750}

\bibitem[\protect\citeauthoryear{Burgh, Nordsieck, Kobulnicky, Williams,
  O'Donoghue, Smith  \& Percival}{Burgh et~al.}{2003}]{burgh_prime_2003}
Burgh E.~B.,  Nordsieck K.~H.,  Kobulnicky H.~A.,  Williams T.~B.,  O'Donoghue
  D.,  Smith M.~P.,   Percival J.~W.,  2003, in Iye M.,  Moorwood A. F.~M.,
  eds,  Society of {Photo}-{Optical} {Instrumentation} {Engineers} ({SPIE})
  {Conference} {Series} Vol. 4841, Instrument {Design} and {Performance} for
  {Optical}/{Infrared} {Ground}-based {Telescopes}. pp 1463--1471,
  \mn@doi{10.1117/12.460312}

\bibitem[\protect\citeauthoryear{Burrows et~al.,}{Burrows
  et~al.}{2005}]{burrows_swift_2005}
Burrows D.~N.,  et~al., 2005, \mn@doi [Space Science Reviews]
  {10.1007/s11214-005-5097-2}, 120, 165

\bibitem[\protect\citeauthoryear{Cannizzaro et~al.,}{Cannizzaro
  et~al.}{2021}]{cannizzaro_accretion_2021}
Cannizzaro G.,  et~al., 2021, \mn@doi [Monthly Notices of the Royal
  Astronomical Society] {10.1093/mnras/stab851}, 504, 792

\bibitem[\protect\citeauthoryear{Cardelli, Clayton  \& Mathis}{Cardelli
  et~al.}{1989}]{cardelli_relationship_1989}
Cardelli J.~A.,  Clayton G.~C.,   Mathis J.~S.,  1989, \mn@doi [The
  Astrophysical Journal] {10.1086/167900}, 345, 245

\bibitem[\protect\citeauthoryear{Carter \& Luminet}{Carter \&
  Luminet}{1983}]{carter_tidal_1983}
Carter B.,  Luminet J.~P.,  1983, Astronomy \& Astrophysics, 121, 97

\bibitem[\protect\citeauthoryear{Cash}{Cash}{1976}]{cash_generation_1976}
Cash W.,  1976, Astronomy and Astrophysics, 52, 307

\bibitem[\protect\citeauthoryear{Cendes, Alexander, Berger, Eftekhari, Williams
   \& Chornock}{Cendes et~al.}{2021}]{cendes_radio_2021}
Cendes Y.,  Alexander K.~D.,  Berger E.,  Eftekhari T.,  Williams P. K.~G.,
  Chornock R.,  2021, \mn@doi [The Astrophysical Journal]
  {10.3847/1538-4357/ac110a}, 919, 127

\bibitem[\protect\citeauthoryear{Cendes et~al.,}{Cendes
  et~al.}{2022}]{cendes_mildly_2022}
Cendes Y.,  et~al., 2022, \mn@doi [The Astrophysical Journal]
  {10.3847/1538-4357/ac88d0}, 938, 28

\bibitem[\protect\citeauthoryear{Chabrier}{Chabrier}{2003}]{chabrier_galactic_2003}
Chabrier G.,  2003, \mn@doi [Publications of the Astronomical Society of the
  Pacific] {10.1086/376392}, 115, 763

\bibitem[\protect\citeauthoryear{Chen, Zaw, Farrar  \& Elgamal}{Chen
  et~al.}{2022}]{chen_uniformly_2022}
Chen Y.-P.,  Zaw I.,  Farrar G.~R.,   Elgamal S.,  2022, \mn@doi [The
  Astrophysical Journal Supplement Series] {10.3847/1538-4365/ac4157}, 258, 29

\bibitem[\protect\citeauthoryear{Conroy \& Gunn}{Conroy \&
  Gunn}{2010}]{conroy_fsps_2010}
Conroy C.,  Gunn J.~E.,  2010, {FSPS}: {Flexible} {Stellar} {Population}
  {Synthesis}

\bibitem[\protect\citeauthoryear{Crawford et~al.,}{Crawford
  et~al.}{2010}]{crawford_pysalt_2010}
Crawford S.~M.,  et~al., 2010, in Silva D.~R.,  Peck A.~B.,   Soifer B.~T.,
  eds,  Society of {Photo}-{Optical} {Instrumentation} {Engineers} ({SPIE})
  {Conference} {Series} Vol. 7737, Observatory {Operations}: {Strategies},
  {Processes}, and {Systems} {III}. p. 773725, \mn@doi{10.1117/12.857000}

\bibitem[\protect\citeauthoryear{Dai, McKinney, Roth, Ramirez-Ruiz  \&
  Miller}{Dai et~al.}{2018}]{dai_unified_2018}
Dai L.,  McKinney J.~C.,  Roth N.,  Ramirez-Ruiz E.,   Miller M.~C.,  2018,
  \mn@doi [The Astrophysical Journal] {10.3847/2041-8213/aab429}, 859, L20

\bibitem[\protect\citeauthoryear{Dey et~al.,}{Dey
  et~al.}{2019}]{dey_overview_2019}
Dey A.,  et~al., 2019, \mn@doi [The Astronomical Journal]
  {10.3847/1538-3881/ab089d}, 157, 168

\bibitem[\protect\citeauthoryear{Engelthaler \& Maksym}{Engelthaler \&
  Maksym}{2023}]{engelthaler_ultraviolet_2023}
Engelthaler E.,  Maksym W.,  2023, in American {Astronomical} {Society}
  {Meeting} {Abstracts}. p. 301.14

\bibitem[\protect\citeauthoryear{Evans et~al.,}{Evans
  et~al.}{2007}]{evans_online_2007}
Evans P.~A.,  et~al., 2007, \mn@doi [Astronomy \& Astrophysics]
  {10.1051/0004-6361:20077530}, 469, 379

\bibitem[\protect\citeauthoryear{Evans et~al.,}{Evans
  et~al.}{2009}]{evans_methods_2009}
Evans P.~A.,  et~al., 2009, \mn@doi [Monthly Notices of the Royal Astronomical
  Society] {10.1111/j.1365-2966.2009.14913.x}, 397, 1177

\bibitem[\protect\citeauthoryear{Flohic, Eracleous, Chartas, Shields  \&
  Moran}{Flohic et~al.}{2006}]{flohic_central_2006}
Flohic H. M. L.~G.,  Eracleous M.,  Chartas G.,  Shields J.~C.,   Moran E.~C.,
  2006, \mn@doi [The Astrophysical Journal] {10.1086/505296}, 647, 140

\bibitem[\protect\citeauthoryear{Foreman-Mackey, Hogg, Lang  \&
  Goodman}{Foreman-Mackey et~al.}{2013}]{foreman-mackey_emcee_2013}
Foreman-Mackey D.,  Hogg D.~W.,  Lang D.,   Goodman J.,  2013, \mn@doi
  [Publications of the Astronomical Society of the Pacific] {10.1086/670067},
  125, 306

\bibitem[\protect\citeauthoryear{Foreman-Mackey, Sick  \&
  Johnson}{Foreman-Mackey et~al.}{2014}]{foreman-mackey_python-fsps_2014}
Foreman-Mackey D.,  Sick J.,   Johnson B.,  2014, python-fsps: {Python}
  bindings to {FSPS} (v0.1.1), \mn@doi{10.5281/zenodo.12157}, \url
  {https://doi.org/10.5281/zenodo.12157}

\bibitem[\protect\citeauthoryear{Franchini, Lodato  \& Facchini}{Franchini
  et~al.}{2016}]{franchini_lensethirring_2016}
Franchini A.,  Lodato G.,   Facchini S.,  2016, \mn@doi [Monthly Notices of the
  Royal Astronomical Society] {10.1093/mnras/stv2417}, 455, 1946

\bibitem[\protect\citeauthoryear{Fruscione et~al.,}{Fruscione
  et~al.}{2006}]{fruscione_ciao_2006}
Fruscione A.,  et~al., 2006, in Silva D.~R.,  Doxsey R.~E.,  eds,  Society of
  {Photo}-{Optical} {Instrumentation} {Engineers} ({SPIE}) {Conference}
  {Series} Vol. 6270, Observatory {Operations}: {Strategies}, {Processes}, and
  {Systems}. SPIE, pp 586 -- 597, \mn@doi{10.1117/12.671760}, \url
  {https://doi.org/10.1117/12.671760}

\bibitem[\protect\citeauthoryear{Fulton, Smith, Moore, Srivastav  \&
  Bruch}{Fulton et~al.}{2022}]{fulton_epessto_2022}
Fulton M.,  Smith K.~W.,  Moore T.,  Srivastav S.,   Bruch R.~J.,  2022,
  Transient Name Server Classification Report, 2022-584, 1

\bibitem[\protect\citeauthoryear{Förster et~al.,}{Förster
  et~al.}{2021}]{forster_automatic_2021}
Förster F.,  et~al., 2021, \mn@doi [The Astronomical Journal]
  {10.3847/1538-3881/abe9bc}, 161, 242

\bibitem[\protect\citeauthoryear{{Gaia Collaboration} et~al.,}{{Gaia
  Collaboration} et~al.}{2021}]{gaia_collaboration_gaia_2021}
{Gaia Collaboration} et~al., 2021, \mn@doi [Astronomy \& Astrophysics]
  {10.1051/0004-6361/202039498}, 649, A6

\bibitem[\protect\citeauthoryear{Gatuzz, García, Kallman, Mendoza  \&
  Gorczyca}{Gatuzz et~al.}{2015}]{gatuzz_ismabs_2015}
Gatuzz E.,  García J.,  Kallman T.~R.,  Mendoza C.,   Gorczyca T.~W.,  2015,
  \mn@doi [The Astrophysical Journal] {10.1088/0004-637X/800/1/29}, 800, 29

\bibitem[\protect\citeauthoryear{Gehrels et~al.,}{Gehrels
  et~al.}{2004}]{gehrels_swift_2004}
Gehrels N.,  et~al., 2004, in {AIP} {Conference} {Proceedings}. pp 637--641,
  \mn@doi{10.1063/1.1810924}, \url {http://arxiv.org/abs/astro-ph/0405233}

\bibitem[\protect\citeauthoryear{Gezari}{Gezari}{2021}]{gezari_tidal_2021}
Gezari S.,  2021, \mn@doi [Annual Review of Astronomy and Astrophysics]
  {10.1146/annurev-astro-111720-030029}, 59, 21

\bibitem[\protect\citeauthoryear{Gezari, Cenko  \& Arcavi}{Gezari
  et~al.}{2017}]{gezari_x-ray_2017}
Gezari S.,  Cenko S.~B.,   Arcavi I.,  2017, \mn@doi [The Astrophysical
  Journal] {10.3847/2041-8213/aaa0c2}, 851, L47

\bibitem[\protect\citeauthoryear{González-Martín, Masegosa, Márquez,
  Guainazzi  \& Jiménez-Bailón}{González-Martín
  et~al.}{2009}]{gonzalez-martin_x-ray_2009}
González-Martín O.,  Masegosa J.,  Márquez I.,  Guainazzi M.,
  Jiménez-Bailón E.,  2009, \mn@doi [Astronomy \& Astrophysics]
  {10.1051/0004-6361/200912288}, 506, 1107

\bibitem[\protect\citeauthoryear{Goodwin et~al.,}{Goodwin
  et~al.}{2022a}]{goodwin_at2019azh_2022}
Goodwin A.~J.,  et~al., 2022a, \mn@doi [Monthly Notices of the Royal
  Astronomical Society] {10.1093/mnras/stac333}, 511, 5328

\bibitem[\protect\citeauthoryear{Goodwin et~al.,}{Goodwin
  et~al.}{2022b}]{goodwin_radio_2022}
Goodwin A.~J.,  et~al., 2022b, \mn@doi [Monthly Notices of the Royal
  Astronomical Society] {10.1093/mnras/stac3127}, 518, 847

\bibitem[\protect\citeauthoryear{Goodwin et~al.,}{Goodwin
  et~al.}{2023}]{goodwin_radio-emitting_2023}
Goodwin A.~J.,  et~al., 2023, \mn@doi [Monthly Notices of the Royal
  Astronomical Society] {10.1093/mnras/stad1258}, 522, 5084

\bibitem[\protect\citeauthoryear{Graham et~al.,}{Graham
  et~al.}{2019}]{graham_zwicky_2019}
Graham M.~J.,  et~al., 2019, \mn@doi [Publications of the Astronomical Society
  of the Pacific] {10.1088/1538-3873/ab006c}, 131, 078001

\bibitem[\protect\citeauthoryear{Greiner, Schwarz, Zharikov  \& Orio}{Greiner
  et~al.}{2000}]{greiner_rx_2000}
Greiner J.,  Schwarz R.,  Zharikov S.,   Orio M.,  2000, Astronomy and
  Astrophysics, 362, L25

\bibitem[\protect\citeauthoryear{Grupe, Thomas  \& Leighly}{Grupe
  et~al.}{1999}]{grupe_rx_1999}
Grupe D.,  Thomas H.~C.,   Leighly K.~M.,  1999, Astronomy and Astrophysics,
  350, L31

\bibitem[\protect\citeauthoryear{Gu \& Cao}{Gu \&
  Cao}{2009}]{gu_anticorrelation_2009}
Gu M.,  Cao X.,  2009, \mn@doi [Monthly Notices of the Royal Astronomical
  Society] {10.1111/j.1365-2966.2009.15277.x}, 399, 349

\bibitem[\protect\citeauthoryear{Guillochon, Ramirez-Ruiz, Rosswog  \&
  Kasen}{Guillochon et~al.}{2009}]{guillochon_three-dimensional_2009}
Guillochon J.,  Ramirez-Ruiz E.,  Rosswog S.,   Kasen D.,  2009, \mn@doi [The
  Astrophysical Journal] {10.1088/0004-637X/705/1/844}, 705, 844

\bibitem[\protect\citeauthoryear{Guillochon, Nicholl, Villar, Mockler, Narayan,
  Mandel, Berger  \& Williams}{Guillochon
  et~al.}{2018}]{guillochon_mosfit_2018}
Guillochon J.,  Nicholl M.,  Villar V.~A.,  Mockler B.,  Narayan G.,  Mandel
  K.~S.,  Berger E.,   Williams P. K.~G.,  2018, \mn@doi [The Astrophysical
  Journal Supplement Series] {10.3847/1538-4365/aab761}, 236, 6

\bibitem[\protect\citeauthoryear{Guo, Shen  \& Wang}{Guo
  et~al.}{2018}]{guo_pyqsofit_2018}
Guo H.,  Shen Y.,   Wang S.,  2018, {PyQSOFit}: {Python} code to fit the
  spectrum of quasars

\bibitem[\protect\citeauthoryear{Hammerstein et~al.,}{Hammerstein
  et~al.}{2023}]{hammerstein_final_2023}
Hammerstein E.,  et~al., 2023, \mn@doi [The Astrophysical Journal]
  {10.3847/1538-4357/aca283}, 942, 9

\bibitem[\protect\citeauthoryear{Higson, Handley, Hobson  \& Lasenby}{Higson
  et~al.}{2019}]{higson_dynamic_2019}
Higson E.,  Handley W.,  Hobson M.,   Lasenby A.,  2019, \mn@doi [Statistics
  and Computing] {10.1007/s11222-018-9844-0}, 29, 891

\bibitem[\protect\citeauthoryear{Holoien et~al.,}{Holoien
  et~al.}{2016}]{holoien_six_2016}
Holoien T. W.-S.,  et~al., 2016, \mn@doi [Monthly Notices of the Royal
  Astronomical Society] {10.1093/mnras/stv2486}, 455, 2918

\bibitem[\protect\citeauthoryear{Holoien et~al.,}{Holoien
  et~al.}{2019}]{holoien_discovery_2019}
Holoien T. W.-S.,  et~al., 2019, \mn@doi [The Astrophysical Journal]
  {10.3847/1538-4357/ab3c66}, 883, 111

\bibitem[\protect\citeauthoryear{Homan et~al.,}{Homan
  et~al.}{2023}]{homan_discovery_2023}
Homan D.,  et~al., 2023, \mn@doi [Astronomy \& Astrophysics]
  {10.1051/0004-6361/202245078}, 672, A167

\bibitem[\protect\citeauthoryear{Horesh, Cenko  \& Arcavi}{Horesh
  et~al.}{2021}]{horesh_delayed_2021}
Horesh A.,  Cenko S.~B.,   Arcavi I.,  2021, \mn@doi [Nature Astronomy]
  {10.1038/s41550-021-01300-8}, 5, 491

\bibitem[\protect\citeauthoryear{Irwin, Henriksen, Krause, Wang, Wiegert,
  Murphy, Heald  \& Perlman}{Irwin et~al.}{2018}]{irwin_erratum_2018}
Irwin J.~A.,  Henriksen R.~N.,  Krause M.,  Wang Q.~D.,  Wiegert T.,  Murphy
  E.~J.,  Heald G.,   Perlman E.,  2018, \mn@doi [The Astrophysical Journal]
  {10.3847/1538-4357/aacaff}, 860, 176

\bibitem[\protect\citeauthoryear{Ivezic}{Ivezic}{2019}]{ivezic_lsst_2019}
Ivezic Z.,  2019, The Astrophysical Journal, 873, 44

\bibitem[\protect\citeauthoryear{Johnson, Leja, Conroy  \& Speagle}{Johnson
  et~al.}{2021}]{johnson_stellar_2021}
Johnson B.~D.,  Leja J.,  Conroy C.,   Speagle J.~S.,  2021, \mn@doi [The
  Astrophysical Journal Supplement Series] {10.3847/1538-4365/abef67}, 254, 22

\bibitem[\protect\citeauthoryear{Jones et~al.,}{Jones
  et~al.}{2009}]{jones_6df_2009}
Jones D.~H.,  et~al., 2009, \mn@doi [Monthly Notices of the Royal Astronomical
  Society] {10.1111/j.1365-2966.2009.15338.x}, 399, 683

\bibitem[\protect\citeauthoryear{Kewley, Dopita, Sutherland, Heisler  \&
  Trevena}{Kewley et~al.}{2001}]{kewley_theoretical_2001}
Kewley L.~J.,  Dopita M.~A.,  Sutherland R.~S.,  Heisler C.~A.,   Trevena J.,
  2001, \mn@doi [The Astrophysical Journal] {10.1086/321545}, 556, 121

\bibitem[\protect\citeauthoryear{Komossa \& Bade}{Komossa \&
  Bade}{1999}]{komossa_giant_1999}
Komossa S.,  Bade N.,  1999, Astronomy and Astrophysics, 343, 775

\bibitem[\protect\citeauthoryear{Komossa \& Greiner}{Komossa \&
  Greiner}{1999}]{komossa_discovery_1999}
Komossa S.,  Greiner J.,  1999, Astronomy and Astrophysics, 349, L45

\bibitem[\protect\citeauthoryear{Komossa et~al.,}{Komossa
  et~al.}{2009}]{komossa_ntt_2009}
Komossa S.,  et~al., 2009, \mn@doi [The Astrophysical Journal]
  {10.1088/0004-637X/701/1/105}, 701, 105

\bibitem[\protect\citeauthoryear{Koposov et~al.,}{Koposov
  et~al.}{2023}]{koposov_joshspeagledynesty_2023}
Koposov S.,  et~al., 2023, joshspeagle/dynesty: v2.1.0,
  \mn@doi{10.5281/zenodo.7600689}, \url
  {https://doi.org/10.5281/zenodo.7600689}

\bibitem[\protect\citeauthoryear{Kraft, Burrows  \& Nousek}{Kraft
  et~al.}{1991}]{kraft_determination_1991}
Kraft R.~P.,  Burrows D.~N.,   Nousek J.~A.,  1991, \mn@doi [The Astrophysical
  Journal] {10.1086/170124}, 374, 344

\bibitem[\protect\citeauthoryear{Lacy et~al.,}{Lacy
  et~al.}{2020}]{lacy_karl_2020}
Lacy M.,  et~al., 2020, \mn@doi [Publications of the Astronomical Society of
  the Pacific] {10.1088/1538-3873/ab63eb}, 132, 035001

\bibitem[\protect\citeauthoryear{Leloudas et~al.,}{Leloudas
  et~al.}{2019}]{leloudas_spectral_2019}
Leloudas G.,  et~al., 2019, \mn@doi [The Astrophysical Journal]
  {10.3847/1538-4357/ab5792}, 887, 218

\bibitem[\protect\citeauthoryear{Liu, Malyali, Rau, Merloni  \& Krumpe}{Liu
  et~al.}{2022}]{liu_srgerosita_2022}
Liu Z.,  Malyali A.,  Rau A.,  Merloni A.,   Krumpe M.,  2022, The Astronomer's
  Telegram, 15259, 1

\bibitem[\protect\citeauthoryear{Liu et~al.,}{Liu
  et~al.}{2023}]{liu_deciphering_2023}
Liu Z.,  et~al., 2023, \mn@doi [Astronomy \& Astrophysics]
  {10.1051/0004-6361/202244805}, 669, A75

\bibitem[\protect\citeauthoryear{Lodato \& Rossi}{Lodato \&
  Rossi}{2011}]{lodato_multiband_2011}
Lodato G.,  Rossi E.~M.,  2011, \mn@doi [Monthly Notices of the Royal
  Astronomical Society] {10.1111/j.1365-2966.2010.17448.x}, 410, 359

\bibitem[\protect\citeauthoryear{Loeb \& Ulmer}{Loeb \&
  Ulmer}{1997}]{loeb_optical_1997}
Loeb A.,  Ulmer A.,  1997, \mn@doi [The Astrophysical Journal]
  {10.1086/304814}, 489, 573

\bibitem[\protect\citeauthoryear{Lu \& Bonnerot}{Lu \&
  Bonnerot}{2020}]{lu_self-intersection_2020}
Lu W.,  Bonnerot C.,  2020, \mn@doi [Monthly Notices of the Royal Astronomical
  Society] {10.1093/mnras/stz3405}, 492, 686

\bibitem[\protect\citeauthoryear{Mainzer et~al.,}{Mainzer
  et~al.}{2014}]{mainzer_initial_2014}
Mainzer A.,  et~al., 2014, \mn@doi [The Astrophysical Journal]
  {10.1088/0004-637X/792/1/30}, 792, 30

\bibitem[\protect\citeauthoryear{Malyali et~al.,}{Malyali
  et~al.}{2021}]{malyali_at_2021}
Malyali A.,  et~al., 2021, \mn@doi [Astronomy \& Astrophysics]
  {10.1051/0004-6361/202039681}, 647, A9

\bibitem[\protect\citeauthoryear{Malyali et~al.,}{Malyali
  et~al.}{2023a}]{malyali_rebrightening_2023}
Malyali A.,  et~al., 2023a, \mn@doi [Monthly Notices of the Royal Astronomical
  Society] {10.1093/mnras/stad022}, 520, 3549

\bibitem[\protect\citeauthoryear{Malyali et~al.,}{Malyali
  et~al.}{2023b}]{malyali_erasst_2023}
Malyali A.,  et~al., 2023b, \mn@doi [Monthly Notices of the Royal Astronomical
  Society] {10.1093/mnras/stad046}, 520, 4209

\bibitem[\protect\citeauthoryear{Masci et~al.,}{Masci
  et~al.}{2019}]{masci_zwicky_2019}
Masci F.~J.,  et~al., 2019, \mn@doi [Publications of the Astronomical Society
  of the Pacific] {10.1088/1538-3873/aae8ac}, 131, 018003

\bibitem[\protect\citeauthoryear{Metzger \& Stone}{Metzger \&
  Stone}{2016}]{metzger_bright_2016}
Metzger B.~D.,  Stone N.~C.,  2016, \mn@doi [Monthly Notices of the Royal
  Astronomical Society] {10.1093/mnras/stw1394}, 461, 948

\bibitem[\protect\citeauthoryear{Miller}{Miller}{2015}]{miller_disk_2015}
Miller M.~C.,  2015, \mn@doi [The Astrophysical Journal]
  {10.1088/0004-637X/805/1/83}, 805, 83

\bibitem[\protect\citeauthoryear{Mockler, Guillochon  \& Ramirez-Ruiz}{Mockler
  et~al.}{2019}]{mockler_weighing_2019}
Mockler B.,  Guillochon J.,   Ramirez-Ruiz E.,  2019, \mn@doi [The
  Astrophysical Journal] {10.3847/1538-4357/ab010f}, 872, 151

\bibitem[\protect\citeauthoryear{Nandra \& Pounds}{Nandra \&
  Pounds}{1994}]{nandra_ginga_1994}
Nandra K.,  Pounds K.~A.,  1994, \mn@doi [Monthly Notices of the Royal
  Astronomical Society] {10.1093/mnras/268.2.405}, 268, 405

\bibitem[\protect\citeauthoryear{Narayan, Mahadevan  \& Quataert}{Narayan
  et~al.}{1998}]{narayan_advection-dominated_1998}
Narayan R.,  Mahadevan R.,   Quataert E.,  1998, in Abramowicz M.~A.,
  Björnsson G.,   Pringle J.~E.,  eds, Theory of {Black} {Hole} {Accretion}
  {Disks}. pp 148--182, \mn@doi{10.48550/arXiv.astro-ph/9803141}

\bibitem[\protect\citeauthoryear{Newville, Stensitzki, Allen  \&
  Ingargiola}{Newville et~al.}{2014}]{newville_lmfit_2014}
Newville M.,  Stensitzki T.,  Allen D.~B.,   Ingargiola A.,  2014, {LMFIT}:
  {Non}-{Linear} {Least}-{Square} {Minimization} and {Curve}-{Fitting} for
  {Python}, \mn@doi{10.5281/zenodo.11813}, \url
  {https://doi.org/10.5281/zenodo.11813}

\bibitem[\protect\citeauthoryear{Nicholl et~al.,}{Nicholl
  et~al.}{2020}]{nicholl_outflow_2020}
Nicholl M.,  et~al., 2020, \mn@doi [Monthly Notices of the Royal Astronomical
  Society] {10.1093/mnras/staa2824}, 499, 482

\bibitem[\protect\citeauthoryear{Piran, Svirski, Krolik, Cheng  \&
  Shiokawa}{Piran et~al.}{2015}]{piran_disk_2015}
Piran T.,  Svirski G.,  Krolik J.,  Cheng R.~M.,   Shiokawa H.,  2015, \mn@doi
  [The Astrophysical Journal] {10.1088/0004-637X/806/2/164}, 806, 164

\bibitem[\protect\citeauthoryear{Predehl et~al.,}{Predehl
  et~al.}{2021}]{predehl_erosita_2021}
Predehl P.,  et~al., 2021, \mn@doi [Astronomy \& Astrophysics]
  {10.1051/0004-6361/202039313}, 647, A1

\bibitem[\protect\citeauthoryear{Rees}{Rees}{1988}]{rees_tidal_1988}
Rees M.~J.,  1988, \mn@doi [Nature] {10.1038/333523a0}, 333, 523

\bibitem[\protect\citeauthoryear{Reines \& Volonteri}{Reines \&
  Volonteri}{2015}]{reines_relations_2015}
Reines A.~E.,  Volonteri M.,  2015, \mn@doi [The Astrophysical Journal]
  {10.1088/0004-637X/813/2/82}, 813, 82

\bibitem[\protect\citeauthoryear{Roming et~al.,}{Roming
  et~al.}{2005}]{roming_swift_2005}
Roming P. W.~A.,  et~al., 2005, \mn@doi [Space Science Reviews]
  {10.1007/s11214-005-5095-4}, 120, 95

\bibitem[\protect\citeauthoryear{Rossi, Stone, Law-Smith, Macleod, Lodato, Dai
  \& Mandel}{Rossi et~al.}{2021}]{rossi_process_2021}
Rossi E.~M.,  Stone N.~C.,  Law-Smith J. A.~P.,  Macleod M.,  Lodato G.,  Dai
  J.~L.,   Mandel I.,  2021, \mn@doi [Space Science Reviews]
  {10.1007/s11214-021-00818-7}, 217, 40

\bibitem[\protect\citeauthoryear{Roth, Kasen, Guillochon  \& Ramirez-Ruiz}{Roth
  et~al.}{2016}]{roth_x-ray_2016}
Roth N.,  Kasen D.,  Guillochon J.,   Ramirez-Ruiz E.,  2016, \mn@doi [The
  Astrophysical Journal] {10.3847/0004-637X/827/1/3}, 827, 3

\bibitem[\protect\citeauthoryear{Sagiv et~al.,}{Sagiv
  et~al.}{2014}]{sagiv_science_2014}
Sagiv I.,  et~al., 2014, \mn@doi [The Astronomical Journal]
  {10.1088/0004-6256/147/4/79}, 147, 79

\bibitem[\protect\citeauthoryear{Saxton, Komossa, Auchettl  \& Jonker}{Saxton
  et~al.}{2020}]{saxton_x-ray_2020}
Saxton R.,  Komossa S.,  Auchettl K.,   Jonker P.~G.,  2020, \mn@doi [Space
  Science Reviews] {10.1007/s11214-020-00708-4}, 216, 85

\bibitem[\protect\citeauthoryear{Saxton, Komossa, Auchettl  \& Jonker}{Saxton
  et~al.}{2021}]{saxton_correction_2021}
Saxton R.,  Komossa S.,  Auchettl K.,   Jonker P.~G.,  2021, \mn@doi [Space
  Science Reviews] {10.1007/s11214-020-00759-7}, 217, 18

\bibitem[\protect\citeauthoryear{Sazonov et~al.,}{Sazonov
  et~al.}{2021}]{sazonov_first_2021}
Sazonov S.,  et~al., 2021, \mn@doi [Monthly Notices of the Royal Astronomical
  Society] {10.1093/mnras/stab2843}, 508, 3820

\bibitem[\protect\citeauthoryear{Schlafly \& Finkbeiner}{Schlafly \&
  Finkbeiner}{2011}]{schlafly_measuring_2011}
Schlafly E.~F.,  Finkbeiner D.~P.,  2011, \mn@doi [The Astrophysical Journal]
  {10.1088/0004-637X/737/2/103}, 737, 103

\bibitem[\protect\citeauthoryear{Shingles et~al.,}{Shingles
  et~al.}{2021}]{shingles_release_2021}
Shingles L.,  et~al., 2021, Transient Name Server AstroNote, 7, 1

\bibitem[\protect\citeauthoryear{Shiokawa, Krolik, Cheng, Piran  \&
  Noble}{Shiokawa et~al.}{2015}]{shiokawa_general_2015}
Shiokawa H.,  Krolik J.~H.,  Cheng R.~M.,  Piran T.,   Noble S.~C.,  2015,
  \mn@doi [The Astrophysical Journal] {10.1088/0004-637X/804/2/85}, 804, 85

\bibitem[\protect\citeauthoryear{Simmonds, Buchner, Salvato, Hsu  \&
  Bauer}{Simmonds et~al.}{2018}]{simmonds_xz_2018}
Simmonds C.,  Buchner J.,  Salvato M.,  Hsu L.-T.,   Bauer F.~E.,  2018,
  \mn@doi [Astronomy \& Astrophysics] {10.1051/0004-6361/201833412}, 618, A66

\bibitem[\protect\citeauthoryear{Skilling}{Skilling}{2004}]{skilling_nested_2004}
Skilling J.,  2004, in Fischer R.,  Preuss R.,   Toussaint U.~V.,  eds,
  American {Institute} of {Physics} {Conference} {Series} Vol. 735, Bayesian
  {Inference} and {Maximum} {Entropy} {Methods} in {Science} and {Engineering}:
  24th {International} {Workshop} on {Bayesian} {Inference} and {Maximum}
  {Entropy} {Methods} in {Science} and {Engineering}. pp 395--405,
  \mn@doi{10.1063/1.1835238}

\bibitem[\protect\citeauthoryear{Skilling}{Skilling}{2006}]{skilling_nested_2006}
Skilling J.,  2006, \mn@doi [Bayesian Analysis] {10.1214/06-BA127}, 1, 833

\bibitem[\protect\citeauthoryear{Smartt et~al.,}{Smartt
  et~al.}{2015}]{smartt_pessto_2015}
Smartt S.~J.,  et~al., 2015, \mn@doi [Astronomy \& Astrophysics]
  {10.1051/0004-6361/201425237}, 579, A40

\bibitem[\protect\citeauthoryear{Smith et~al.,}{Smith
  et~al.}{2020}]{smith_design_2020}
Smith K.~W.,  et~al., 2020, \mn@doi [Publications of the Astronomical Society
  of the Pacific] {10.1088/1538-3873/ab936e}, 132, 085002

\bibitem[\protect\citeauthoryear{Speagle}{Speagle}{2020}]{speagle_dynesty_2020}
Speagle J.~S.,  2020, \mn@doi [Monthly Notices of the Royal Astronomical
  Society] {10.1093/mnras/staa278}, 493, 3132

\bibitem[\protect\citeauthoryear{Stanek \& Kochanek}{Stanek \&
  Kochanek}{2022}]{stanek_asas-sn_2022}
Stanek K.~Z.,  Kochanek C.~S.,  2022, Transient Name Server Discovery Report,
  2022-559, 1

\bibitem[\protect\citeauthoryear{Stern et~al.,}{Stern
  et~al.}{2012}]{stern_mid-infrared_2012}
Stern D.,  et~al., 2012, \mn@doi [The Astrophysical Journal]
  {10.1088/0004-637X/753/1/30}, 753, 30

\bibitem[\protect\citeauthoryear{Stone \& Loeb}{Stone \&
  Loeb}{2012}]{stone_observing_2012}
Stone N.,  Loeb A.,  2012, \mn@doi [Physical Review Letters]
  {10.1103/PhysRevLett.108.061302}, 108, 061302

\bibitem[\protect\citeauthoryear{Stone, Sari  \& Loeb}{Stone
  et~al.}{2013}]{stone_consequences_2013}
Stone N.,  Sari R.,   Loeb A.,  2013, \mn@doi [Monthly Notices of the Royal
  Astronomical Society] {10.1093/mnras/stt1270}, 435, 1809

\bibitem[\protect\citeauthoryear{Sunyaev et~al.,}{Sunyaev
  et~al.}{2021}]{sunyaev_srg_2021}
Sunyaev R.,  et~al., 2021, \mn@doi [Astronomy \& Astrophysics]
  {10.1051/0004-6361/202141179}, 656, A132

\bibitem[\protect\citeauthoryear{{The CASA Team} et~al.,}{{The CASA Team}
  et~al.}{2022}]{the_casa_team_casa_2022}
{The CASA Team} et~al., 2022, \mn@doi [Publications of the Astronomical Society
  of the Pacific] {10.1088/1538-3873/ac9642}, 134, 114501

\bibitem[\protect\citeauthoryear{Tody}{Tody}{1986}]{tody_iraf_1986}
Tody D.,  1986, in Crawford D.~L.,  ed.,  Society of {Photo}-{Optical}
  {Instrumentation} {Engineers} ({SPIE}) {Conference} {Series} Vol. 627,
  Instrumentation in astronomy {VI}. p.~733, \mn@doi{10.1117/12.968154}

\bibitem[\protect\citeauthoryear{Tonry et~al.,}{Tonry
  et~al.}{2018}]{tonry_atlas_2018}
Tonry J.~L.,  et~al., 2018, \mn@doi [Publications of the Astronomical Society
  of the Pacific] {10.1088/1538-3873/aabadf}, 130, 064505

\bibitem[\protect\citeauthoryear{Trümper}{Trümper}{1982}]{trumper_rosat_1982}
Trümper J.,  1982, \mn@doi [Advances in Space Research]
  {https://doi.org/10.1016/0273-1177(82)90070-9}, 2, 241

\bibitem[\protect\citeauthoryear{Verner, Ferland, Korista  \& Yakovlev}{Verner
  et~al.}{1996}]{verner_atomic_1996}
Verner D.~A.,  Ferland G.~J.,  Korista K.~T.,   Yakovlev D.~G.,  1996, \mn@doi
  [The Astrophysical Journal] {10.1086/177435}, 465, 487

\bibitem[\protect\citeauthoryear{Wang, Zhou, Wang, Lu  \& Xu}{Wang
  et~al.}{2011}]{wang_transient_2011}
Wang T.-G.,  Zhou H.-Y.,  Wang L.-F.,  Lu H.-L.,   Xu D.,  2011, \mn@doi [The
  Astrophysical Journal] {10.1088/0004-637X/740/2/85}, 740, 85

\bibitem[\protect\citeauthoryear{Wang, Zhou, Komossa, Wang, Yuan  \& Yang}{Wang
  et~al.}{2012}]{wang_extreme_2012}
Wang T.-G.,  Zhou H.-Y.,  Komossa S.,  Wang H.-Y.,  Yuan W.,   Yang C.,  2012,
  \mn@doi [The Astrophysical Journal] {10.1088/0004-637X/749/2/115}, 749, 115

\bibitem[\protect\citeauthoryear{Wevers et~al.,}{Wevers
  et~al.}{2022}]{wevers_elliptical_2022}
Wevers T.,  et~al., 2022, \mn@doi [Astronomy \& Astrophysics]
  {10.1051/0004-6361/202142616}, 666, A6

\bibitem[\protect\citeauthoryear{Willingale, Starling, Beardmore, Tanvir  \&
  O'Brien}{Willingale et~al.}{2013}]{willingale_calibration_2013}
Willingale R.,  Starling R. L.~C.,  Beardmore A.~P.,  Tanvir N.~R.,   O'Brien
  P.~T.,  2013, \mn@doi [Monthly Notices of the Royal Astronomical Society]
  {10.1093/mnras/stt175}, 431, 394

\bibitem[\protect\citeauthoryear{Wilms, Allen  \& McCray}{Wilms
  et~al.}{2000}]{wilms_absorption_2000}
Wilms J.,  Allen A.,   McCray R.,  2000, \mn@doi [The Astrophysical Journal]
  {10.1086/317016}, 542, 914

\bibitem[\protect\citeauthoryear{Wright et~al.,}{Wright
  et~al.}{2010}]{wright_wide-field_2010}
Wright E.~L.,  et~al., 2010, \mn@doi [The Astronomical Journal]
  {10.1088/0004-6256/140/6/1868}, 140, 1868

\bibitem[\protect\citeauthoryear{Yalinewich, Guillochon, Sari  \&
  Loeb}{Yalinewich et~al.}{2019}]{yalinewich_shock_2019}
Yalinewich A.,  Guillochon J.,  Sari R.,   Loeb A.,  2019, \mn@doi [Monthly
  Notices of the Royal Astronomical Society] {10.1093/mnras/sty2809}, 482, 2872

\bibitem[\protect\citeauthoryear{Yuan et~al.,}{Yuan
  et~al.}{2018}]{yuan_einstein_2018}
Yuan W.,  et~al., 2018, in den Herder J.-W.~A.,  Nakazawa K.,   Nikzad S.,
  eds, Space {Telescopes} and {Instrumentation} 2018: {Ultraviolet} to {Gamma}
  {Ray}. SPIE, Austin, United States, p.~76, \mn@doi{10.1117/12.2313358}, \url
  {https://www.spiedigitallibrary.org/conference-proceedings-of-spie/10699/2313358/Einstein-Probe--a-lobster-eye-telescope-for-monitoring-the/10.1117/12.2313358.full}

\bibitem[\protect\citeauthoryear{van Velzen et~al.,}{van Velzen
  et~al.}{2019}]{van_velzen_first_2019}
van Velzen S.,  et~al., 2019, \mn@doi [The Astrophysical Journal]
  {10.3847/1538-4357/aafe0c}, 872, 198

\bibitem[\protect\citeauthoryear{van Velzen, Holoien, Onori, Hung  \&
  Arcavi}{van Velzen et~al.}{2020}]{van_velzen_optical-ultraviolet_2020}
van Velzen S.,  Holoien T. W.-S.,  Onori F.,  Hung T.,   Arcavi I.,  2020,
  \mn@doi [Space Science Reviews] {10.1007/s11214-020-00753-z}, 216, 124

\bibitem[\protect\citeauthoryear{van Velzen et~al.,}{van Velzen
  et~al.}{2021}]{van_velzen_seventeen_2021}
van Velzen S.,  et~al., 2021, \mn@doi [The Astrophysical Journal]
  {10.3847/1538-4357/abc258}, 908, 4

\makeatother
\end{thebibliography}

%\begin{thebibliography}{99}
%\end{thebibliography}

%%%%%%%%%%%%%%%%%%%%%%%%%%%%%%%%%%%%%%%%%%%%%%%%%%

%%%%%%%%%%%%%%%%% APPENDICES %%%%%%%%%%%%%%%%%%%%%

\appendix
\section{Additional photometric information}
The optical and UV photometry of \dsb \, is presented in Table~\ref{tab:uvot_photometry}.
\begin{table}
\centering
\caption{Optical and UV photometry of \dsb, corrected for Galactic extinction, and with 3$\sigma$ upper limits presented.}
\label{tab:uvot_photometry}
\begin{tabular}{cccc}
\hline
MJD & Instrument & Filter & Magnitude \\
\hline
59601.660 & ATLAS & o & $<$19.21 \\
59610.660 & ATLAS & o & $<$18.97 \\
59619.140 & ATLAS & c & $<$18.14 \\
59624.095 & ATLAS & o & $<$18.69 \\
59624.118 & ATLAS & o & $<$19.59 \\
59627.077 & ATLAS & o & $<$18.90 \\
59627.082 & ATLAS & o & $<$18.90 \\
59627.086 & ATLAS & o & $<$19.01 \\
59627.092 & ATLAS & o & $<$18.97 \\
59628.113 & ATLAS & o & $<$18.84 \\
59628.113 & ATLAS & o & 18.88 $\pm$ 0.34 \\
59628.127 & ATLAS & o & $<$19.13 \\
59628.129 & ATLAS & o & $<$18.97 \\
59628.129 & ATLAS & o & 18.64 $\pm$ 0.26 \\
59628.139 & ATLAS & o & $<$18.58 \\
59628.140 & ATLAS & o & $<$18.46 \\
59629.365 & ATLAS & o & 18.45 $\pm$ 0.13 \\
59630.344 & ATLAS & o & 18.19 $\pm$ 0.13 \\
59630.347 & ATLAS & o & 18.83 $\pm$ 0.20 \\
59630.364 & ATLAS & o & 17.73 $\pm$ 0.09 \\
59630.371 & ATLAS & o & 18.06 $\pm$ 0.12 \\
59637.330 & ATLAS & o & 17.77 $\pm$ 0.04 \\
59643.113 & \textit{Swift} & UVW1 & 16.24 $\pm$ 0.05 \\
59643.114 & \textit{Swift} & U & 16.24 $\pm$ 0.06 \\
59643.114 & \textit{Swift} & B & 15.68 $\pm$ 0.06 \\
59643.116 & \textit{Swift} & UVW2 & 15.52 $\pm$ 0.05 \\
59643.118 & \textit{Swift} & V & 15.37 $\pm$ 0.07 \\
59643.120 & \textit{Swift} & UVM2 & 15.87 $\pm$ 0.05 \\
59644.110 & ATLAS & o & 18.12 $\pm$ 0.05 \\
59645.340 & ATLAS & c & 17.82 $\pm$ 0.06 \\
59647.070 & ATLAS & o & 17.77 $\pm$ 0.05 \\
59648.070 & ATLAS & o & 17.85 $\pm$ 0.03 \\
59649.769 & \textit{Swift} & UVW1 & 16.28 $\pm$ 0.06 \\
59649.771 & \textit{Swift} & U & 16.16 $\pm$ 0.06 \\
59649.771 & \textit{Swift} & B & 15.64 $\pm$ 0.06 \\
59649.773 & \textit{Swift} & UVW2 & 15.61 $\pm$ 0.05 \\
59649.774 & \textit{Swift} & V & 15.34 $\pm$ 0.07 \\
59649.776 & \textit{Swift} & UVM2 & 15.91 $\pm$ 0.05 \\
59653.330 & ATLAS & o & 17.90 $\pm$ 0.04 \\
59654.350 & ATLAS & o & 18.27 $\pm$ 0.06 \\
59655.400 & ATLAS & o & 18.20 $\pm$ 0.08 \\
59656.020 & ATLAS & o & 18.26 $\pm$ 0.09 \\
59656.152 & \textit{Swift} & UVW1 & 16.94 $\pm$ 0.08 \\
59656.153 & \textit{Swift} & U & 16.60 $\pm$ 0.09 \\
59656.154 & \textit{Swift} & B & 16.04 $\pm$ 0.08 \\
59656.156 & \textit{Swift} & UVW2 & 16.54 $\pm$ 0.07 \\
59656.158 & \textit{Swift} & V & 15.29 $\pm$ 0.08 \\
59656.160 & \textit{Swift} & UVM2 & 16.59 $\pm$ 0.08 \\
59657.330 & ATLAS & o & 18.29 $\pm$ 0.08 \\
59658.300 & ATLAS & o & 18.23 $\pm$ 0.09 \\
59662.600 & ATLAS & o & 18.74 $\pm$ 0.25 \\
59663.970 & \textit{Swift} & UVW1 & 17.26 $\pm$ 0.12 \\
59666.160 & ATLAS & o & 19.01 $\pm$ 0.12 \\
59668.126 & \textit{Swift} & UVW1 & 17.42 $\pm$ 0.09 \\
59668.127 & \textit{Swift} & U & 16.98 $\pm$ 0.09 \\
59668.127 & \textit{Swift} & B & 16.12 $\pm$ 0.07 \\
59668.128 & \textit{Swift} & UVW2 & 16.85 $\pm$ 0.07 \\
59668.129 & \textit{Swift} & V & 15.58 $\pm$ 0.08 \\
59668.130 & \textit{Swift} & UVM2 & 17.09 $\pm$ 0.10 \\
59669.190 & ATLAS & o & 19.50 $\pm$ 0.15 \\
59670.310 & ATLAS & o & 19.94 $\pm$ 0.23 \\
59670.461 & ZTF & r & 19.10 $\pm$ 0.11 \\
59670.484 & ZTF & g & 19.07 $\pm$ 0.07 \\
59673.030 & ATLAS & c & 18.51 $\pm$ 0.10 \\
59673.417 & ZTF & g & 19.08 $\pm$ 0.11 \\
59673.481 & ZTF & r & 19.18 $\pm$ 0.17 \\
59673.807 & \textit{Swift} & UVW1 & 17.58 $\pm$ 0.11 \\
59673.809 & \textit{Swift} & U & 16.99 $\pm$ 0.10 \\
59673.809 & \textit{Swift} & B & 16.11 $\pm$ 0.08 \\
59673.811 & \textit{Swift} & UVW2 & 17.17 $\pm$ 0.12 \\
59674.260 & ATLAS & o & 19.33 $\pm$ 0.12 \\
59675.461 & ZTF & g & 19.15 $\pm$ 0.10 \\
59677.314 & \textit{Swift} & UVW1 & 17.84 $\pm$ 0.11 \\
59677.315 & \textit{Swift} & U & 17.19 $\pm$ 0.10 \\
59677.316 & \textit{Swift} & B & 16.15 $\pm$ 0.07 \\
59677.317 & \textit{Swift} & UVW2 & 17.26 $\pm$ 0.10 \\
59677.360 & ATLAS & o & 19.04 $\pm$ 0.22 \\
59677.462 & ZTF & g & 19.18 $\pm$ 0.14 \\
59678.280 & ATLAS & o & 19.13 $\pm$ 0.12 \\
59680.020 & ATLAS & o & 19.94 $\pm$ 0.17 \\
59682.270 & ATLAS & o & $<$20.34 \\
59683.040 & ATLAS & o & $<$17.64 \\
59683.207 & \textit{Swift} & UVW1 & 17.81 $\pm$ 0.16 \\
59683.208 & \textit{Swift} & U & 17.34 $\pm$ 0.16 \\
59683.208 & \textit{Swift} & B & 16.15 $\pm$ 0.10 \\
59683.210 & \textit{Swift} & UVW2 & 17.58 $\pm$ 0.15 \\
59683.211 & \textit{Swift} & V & 15.59 $\pm$ 0.12 \\
59683.213 & \textit{Swift} & UVM2 & 17.47 $\pm$ 0.13 \\
59683.440 & ZTF & g & 19.55 $\pm$ 0.26 \\
59684.416 & ZTF & r & $<$19.43 \\
59684.451 & ZTF & g & 19.38 $\pm$ 0.28 \\
59685.190 & ATLAS & o & 19.65 $\pm$ 0.34 \\
59689.418 & ZTF & r & $<$19.06 \\
59689.461 & ZTF & g & $<$19.06 \\
59690.320 & ATLAS & o & 19.38 $\pm$ 0.24 \\
59693.280 & ATLAS & o & $<$20.41 \\
59694.330 & ATLAS & o & $<$20.21 \\
59694.377 & ZTF & g & $<$20.38 \\
59696.337 & ZTF & g & 19.92 $\pm$ 0.21 \\
59696.447 & ZTF & r & 20.04 $\pm$ 0.30 \\
59696.480 & \textit{Swift} & UVM2 & 17.74 $\pm$ 0.16 \\
59696.482 & \textit{Swift} & UVW1 & 18.24 $\pm$ 0.17 \\
59696.482 & \textit{Swift} & U & 17.34 $\pm$ 0.11 \\
59696.483 & \textit{Swift} & UVW2 & 18.15 $\pm$ 0.18 \\
59697.300 & ATLAS & o & 19.55 $\pm$ 0.14 \\
59698.270 & ATLAS & o & 19.62 $\pm$ 0.17 \\
59699.020 & ATLAS & o & 20.06 $\pm$ 0.22 \\
59699.355 & ZTF & r & $<$20.37 \\
59699.440 & ZTF & g & 19.80 $\pm$ 0.17 \\
59700.070 & ATLAS & o & $<$20.69 \\
59701.270 & ATLAS & o & $<$18.80 \\
59701.376 & ZTF & g & 20.46 $\pm$ 0.31 \\
59701.418 & ZTF & g & 19.89 $\pm$ 0.17 \\
59701.424 & \textit{Swift} & UVM2 & 18.61 $\pm$ 0.27 \\
59701.425 & \textit{Swift} & UVW1 & 18.17 $\pm$ 0.16 \\
59701.426 & \textit{Swift} & U & 17.41 $\pm$ 0.11 \\
59701.427 & \textit{Swift} & UVW2 & 17.76 $\pm$ 0.13 \\
59702.290 & ATLAS & o & $<$20.37 \\
59703.470 & ATLAS & o & $<$20.87 \\
59704.748 & \textit{Swift} & UVM2 & 17.90 $\pm$ 0.14 \\
59704.750 & \textit{Swift} & UVW1 & 18.16 $\pm$ 0.13 \\
59704.752 & \textit{Swift} & U & 17.48 $\pm$ 0.09 \\
59704.753 & \textit{Swift} & UVW2 & 18.19 $\pm$ 0.15 \\
59705.310 & ATLAS & o & $<$18.27 \\
59706.240 & ATLAS & o & $<$20.64 \\
59707.397 & ZTF & r & $<$20.14 \\
59709.230 & ATLAS & o & 19.69 $\pm$ 0.21 \\
59709.378 & ZTF & r & $<$20.13 \\
59710.310 & ATLAS & o & $<$17.74 \\
59712.540 & ATLAS & o & $<$18.81 \\
59723.990 & ATLAS & o & 19.78 $\pm$ 0.25 \\
59725.200 & ATLAS & o & 19.66 $\pm$ 0.12 \\
59726.590 & ATLAS & o & 19.85 $\pm$ 0.15 \\
59729.054 & \textit{Swift} & U & 17.24 $\pm$ 0.16 \\
59729.055 & \textit{Swift} & UVW2 & 18.19 $\pm$ 0.24 \\
59729.577 & \textit{Swift} & UVW1 & 18.15 $\pm$ 0.21 \\
59729.978 & \textit{Swift} & UVM2 & 18.38 $\pm$ 0.20 \\
59730.920 & ATLAS & o & $<$20.52 \\
59731.960 & ATLAS & o & 20.09 $\pm$ 0.24 \\
59734.960 & ATLAS & o & 20.40 $\pm$ 0.36 \\
59735.804 & \textit{Swift} & UVM2 & 18.46 $\pm$ 0.22 \\
59735.808 & \textit{Swift} & UVW1 & 18.52 $\pm$ 0.17 \\
59735.810 & \textit{Swift} & U & 17.47 $\pm$ 0.10 \\
59735.814 & \textit{Swift} & UVW2 & 18.37 $\pm$ 0.15 \\
59735.910 & ATLAS & o & 19.72 $\pm$ 0.14 \\
59737.180 & ATLAS & o & $<$20.72 \\
59745.310 & ATLAS & o & $<$18.20 \\
59749.140 & ATLAS & o & $<$20.22 \\
59750.190 & ATLAS & o & $<$19.89 \\
59765.100 & ATLAS & o & $<$19.91 \\
59766.790 & ATLAS & o & $<$20.59 \\
59767.820 & ATLAS & o & $<$20.07 \\
59773.150 & ATLAS & o & 19.15 $\pm$ 0.34 \\
59775.840 & ATLAS & o & $<$20.29 \\
59781.070 & ATLAS & o & 19.45 $\pm$ 0.18 \\
59782.100 & ATLAS & o & $<$20.57 \\
59783.840 & ATLAS & c & $<$20.43 \\
59785.100 & ATLAS & o & $<$20.55 \\
59786.060 & ATLAS & o & $<$20.03 \\
59787.860 & ATLAS & c & 20.58 $\pm$ 0.36 \\
59789.060 & ATLAS & c & 20.40 $\pm$ 0.28 \\
59790.410 & ATLAS & c & 20.53 $\pm$ 0.29 \\
59791.780 & ATLAS & c & $<$20.71 \\
59793.080 & ATLAS & c & 20.31 $\pm$ 0.27 \\
59795.830 & ATLAS & o & $<$19.88 \\
59799.810 & ATLAS & o & $<$16.37 \\
59801.080 & ATLAS & o & $<$19.45 \\
59802.070 & ATLAS & o & $<$17.68 \\
59805.020 & ATLAS & o & $<$20.24 \\
59806.450 & ATLAS & o & $<$20.44 \\
59807.750 & ATLAS & o & $<$20.52 \\
59814.790 & ATLAS & o & $<$20.43 \\
59815.740 & ATLAS & o & 19.70 $\pm$ 0.35 \\
59817.520 & ATLAS & c & 19.96 $\pm$ 0.15 \\
59818.010 & ATLAS & c & $<$19.87 \\
59818.750 & ATLAS & o & $<$20.19 \\
59820.990 & ATLAS & c & 19.22 $\pm$ 0.17 \\
59821.010 & ATLAS & c & 20.30 $\pm$ 0.32 \\
59822.050 & ATLAS & c & 20.03 $\pm$ 0.25 \\
59822.790 & ATLAS & o & $<$18.97 \\
59825.980 & ATLAS & o & $<$19.02 \\
59828.990 & ATLAS & o & $<$19.25 \\
59830.770 & ATLAS & o & $<$19.78 \\
59855.514 & \textit{Swift} & UVW2 & 18.53 $\pm$ 0.15 \\
59855.518 & \textit{Swift} & UVM2 & 18.52 $\pm$ 0.17 \\
59855.524 & \textit{Swift} & UVW1 & 18.61 $\pm$ 0.13 \\
\end{tabular}
\end{table}

\section{Additional spectroscopic information}\label{sec:appendix_spec_reduction}
\begin{figure}
    \centering
    \includegraphics[scale=0.8]{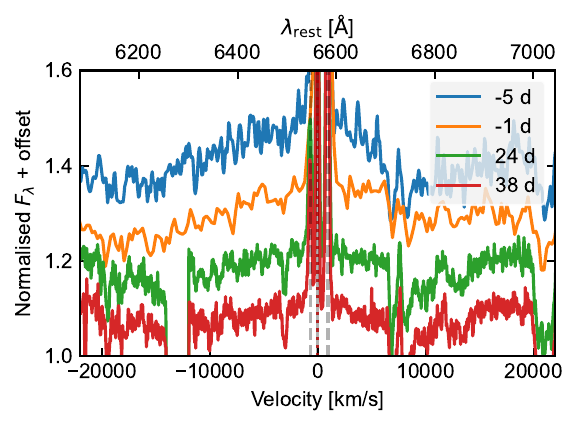}
    \includegraphics[scale=0.8]{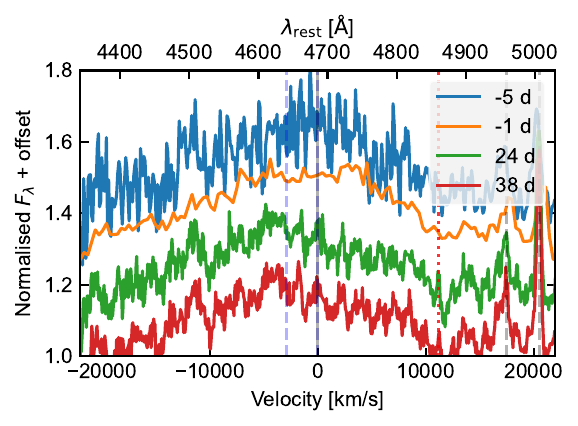}
    \caption{Evolution of the H$\alpha$ (top) and \ion{He}{II} (bottom) line regions for the first four spectra that clearly show broad emission lines. The vertical lines mark [\ion{N}{II}]~6548{\AA}, H$\alpha$, and [\ion{N}{II}]~6583{\AA} in the top panel (left to right), and \ion{N}{III}~4640{\AA}, \ion{He}{II}~4686{\AA}, H$\beta$, [\ion{O}{III}] 4959{\AA}  and [\ion{O}{III}] 5007{\AA} in the bottom panel.}
    \label{fig:optical_spectroscopic_zoomins}
\end{figure}

\begin{figure}
    \centering
    \includegraphics[scale=0.8]{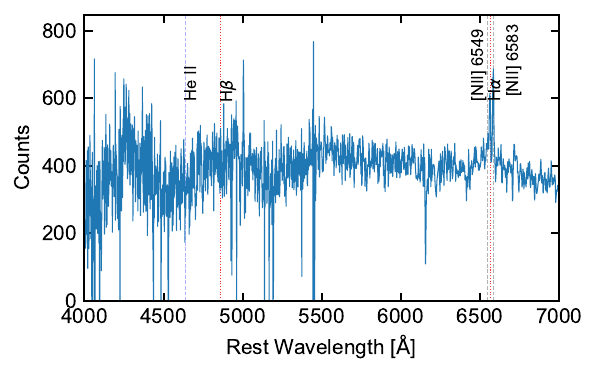}
    \caption{Archival optical spectrum of the host galaxy of \dsb \, taken in 2002 by 6dFGS \citep{jones_6df_2009}. The broad \ion{He}{II} emission complex around 4640~\AA seen in the optical specroscopic follow-up campaign of \dsb \, is clearly not detected. The publicly available 6dFGS spectra do not have absolute flux calibrations.}
    \label{fig:archival_optical_spectrum}
\end{figure}

A zoom-in on the evolution of the broad \ion{He}{II} and H$\alpha$ line profiles is presented in Fig.~\ref{fig:optical_spectroscopic_zoomins}, whilst details on the spectroscopic observations and data reduction are presented below (see Table~\ref{tab:log_spectroscopy} for an observation log). The archival 6dFGS optical spectrum taken in 2002 is plotted in Fig.~\ref{fig:archival_optical_spectrum}.
%[\ion{N}{II}]~6548{\AA} and 6583~{\AA}, and the high-ionisation lines [\ion{O}{III}] 4959{\AA} and 5007~{\AA})
\begin{table}
\centering
\caption{Spectroscopic observations of \dsb.}
\label{tab:log_spectroscopy}
\begin{tabular}{cccccc}
\hline
MJD & UT & Telescope & Instrument & Airmass & Exposure [s] \\
\hline
59636 & 2022-02-26 & LCO FTN & FLOYDS & 1.4 & 2400 \\
59640 & 2022-03-02 & NTT & EFOSC2 & 1.0 & 600 \\
59665 & 2022-03-27 & SALT & RSS & 1.3 & 500 \\
59680 & 2022-04-11 & SALT & RSS & 1.2 & 500 \\
59680 & 2022-04-11 & LCO FTS & FLOYDS & 1.1 & 2400 \\
59699 & 2022-04-30 & LCO FTS & FLOYDS & 1.2 & 2400 \\
59726 & 2022-05-27 & LCO FTN & FLOYDS & 1.5 & 2400 \\
59767 & 2022-07-07 & NTT & EFOSC2 & 1.2 & 1800 \\
59781 & 2022-07-21 & LCO FTS & FLOYDS & 1.4 & 2400 \\
\end{tabular}
\end{table}

\textit{FLOYDS}: \dsb\ was observed five times with the FLOYDS spectrographs mounted at the Las Cumbres Observatory 2\,m telescopes at Siding Springs Observatory (FTS, Australia) and Haleakala (FTN, Hawaii). Observations were performed on 2022-2-26 (FTN), 2022-4-11 and 2022-4-30 (both FTS), 2022-5-27 (FTN), and 2022-7-21 (FTS) with 2400\,s exposure each. The two identical FLOYDS are cross-dispersed, low-resolution spectrographs covering the wavelength range from 320 to 1000\,nm with a resolution between R=400 and R=700. Data were automatically processed by the standard FLOYDS pipeline.

\textit{SALT}: Spectroscopy of \dsb\ was undertaken with the Southern African Large Telescope (SALT; \citealt{buckley_completion_2006}) using the Robert Stobie Spectrograph (RSS, \citealt{burgh_prime_2003}), on two nights: 2022 March 27 and 2022 April 11, starting 23:02 and 03:28 UTC, respectively.  Two consecutive 500 sec exposure spectra with the PG0900 grating were obtained on each night. The spectra covered the region 4345$-$7400\AA at a mean resolution of 5.7\AA. The spectra were reduced using the PyRAF-based PySALT package (\citealt{crawford_pysalt_2010})\footnote{\url{https://astronomers.salt.ac.za/software/}}, which includes corrections for gain and cross-talk, and performs bias subtraction. We extracted the science spectrum using standard IRAF\footnote{https://iraf-community.github.io/} tasks, including wavelength calibration (Xenon calibration lamp exposures were taken, immediately after the science spectra), background subtraction, and 1D spectra extraction. Due to SALT's optical design, absolute flux calibration is not possible\footnote{The entrance pupil changes its position during an observation, resulting in a changing effective collecting area.}.
Observations of spectrophotometric standards during twilight were used to obtain relative flux calibration.
 
\textit{EFOSC2}: 
The source was re-observed with the EFOSC2 spectrograph mounted at the ESO New Technology Telescope (NTT) at La Silla observatory on 2022-7-7 as part of a program targeting eROSITA-selected TDEs (PI: Grotova; 109.23JL.001). Observations were performed with Grism\#13 covering approx. 3600-9300\,\AA\ with a resolution of $\approx21$\,\AA, and an exposure time of 1800s. The data were standard reduced with the IRAF \citep{tody_iraf_1986} community distribution\footnote{\url{https://iraf-community.github.io}}. The EFOSC2 spectrum taken on 2022-03-02 also used Grism\#13, and was reduced by the ePESSTO+ team \citep{smartt_pessto_2015} using version \texttt{ntt\_2.4.0} of their NTT reduction pipeline\footnote{\url{https://github.com/svalenti/pessto}}; the reduced spectrum was downloaded from the Weizmann Interactive Supernova Data Repository (WISeREP\footnote{\url{https://www.wiserep.org/}}).

\bsp	% typesetting comment
\label{lastpage}
\end{document}